# Deep-Learning based Inverse Modeling Approaches: A Subsurface Flow Example

Nanzhe Wang[a], Haibin Chang[a,*], and Dongxiao Zhang[b,*]

[a] BIC-ESAT, ERE, and SKLTCS, College of Engineering, Peking University, Beijing 100871, P. R. China
[b] School of Environmental Science and Engineering, Southern University of Science and Technology, Shenzhen 518055, P. R. China

[*] Corresponding author: E-mail address: changhaibin@pku.edu.cn (Haibin Chang); zhangdx@sustech.edu.cn (Dongxiao Zhang)

**Key Points:**

- Two categories of innovative deep-learning based inverse modeling methods are proposed and compared.

- The deep-learning surrogate-based inversion methods can accelerate the inversion process significantly.

- The direct-deep-learning-inversion method works well in cases with sparse spatial measurements or imprecise prior statistics.

**Abstract**

Deep-learning has achieved good performance and shown great potential for solving forward and inverse problems. In this work, two categories of innovative deep-learning based inverse modeling methods are proposed and compared. The first category is deep-learning surrogate-based inversion methods, in which the Theory-guided Neural Network (TgNN) is constructed as a deep-learning surrogate for problems with uncertain model parameters. By incorporating physical laws and other constraints, the TgNN surrogate can be constructed with limited simulation runs and accelerate the inversion process significantly. Three TgNN surrogate-based inversion methods are proposed, including the gradient method, the iterative ensemble smoother (IES), and the training method. The second category is direct-deep-learning-inversion methods, in which TgNN constrained with geostatistical information, named TgNN-geo, is proposed for direct inverse modeling. In TgNN-geo, two neural networks are introduced to approximate the respective random model parameters and the solution. Since the prior geostatistical information can be incorporated, the direct-inversion method based on TgNN-geo works well, even in cases with sparse spatial measurements or imprecise prior statistics. Although the proposed deep-learning based inverse modeling methods are general in nature, and thus applicable to a wide variety of problems, they are tested with several subsurface flow problems. It is found that satisfactory results are obtained with a high efficiency. Moreover, both the advantages and disadvantages are further analyzed for the proposed two categories of deep-learning based inversion methods.

**Keywords**: Deep-learning; Inverse modeling; Theory-guided Neural Network; Subsurface flow.

# 1 Introduction

Inverse modeling aims to infer uncertain parameters of a system with noisy observations of the system response, which has been widely utilized in various scientific and engineering practices, such as seismic inversion (Bunks et al., 1995), petroleum reservoir history matching (Oliver et al., 2008), aquifer parameter estimation in hydrology (Carrera & Neuman, 1986a, b, c), and medical imaging (Arridge, 1999). Under certain conditions, the inverse problems can be viewed as optimization problems, in which the model parameters are modified, such that the predictions from forward models can match the measurements. In addition, prior knowledge can be used to regularize the objective function and to obtain the maximum *a posteriori* (MAP) estimate of the model parameters.

The gradient-based method is a straightforward and common way to perform inverse modeling tasks. Anterion et al. (1989) presented a rigorous analytical method to calculate the gradients of observations with respect to reservoir characteristics, which can then be used to assist the process of parameter adjustment for history matching. For calculating the required gradients in inverse modeling, the adjoint method is a frequently utilized technique (Chavent et al., 1975; Chen et al., 1974; Wasserman et al., 1975; Yang & Watson, 1988). Carrera and Neuman (1986a, b, c) estimated the aquifer model parameters with maximum likelihood estimation theory, in which the adjoint method was adopted to calculate the gradients of minimization criterions with respect to the model parameters. Wu et al. (1999) used the adjoint method to calculate the sensitivity coefficients of wellbore pressures and water-oil-ratios to model parameters, and then the Gauss-Newton method was employed to optimize the objective function of history matching for two-phase problems. Li et al. (2003) extended the adjoint method to calculate the sensitivity coefficients for history matching of three-dimensional, three-phase flow problems. Although the adjoint method achieves high efficiency, which only requires one forward simulation to calculate the sensitivity, it is tedious to construct adjoint equations, especially for complex problems.

Different from gradient-based methods, ensemble-based inverse modeling methods have attained numerous successes due to their ease of implementation and capability of dealing with large-scale problems. The ensemble Kalman filter (EnKF), proposed by Evensen (1994), is an effective ensemble-based method for inverse modeling, and has been extensively used in various fields, such as meteorology (Houtekamer & Mitchell, 2001; Houtekamer et al., 2005), petroleum engineering (Chang et al., 2010; Gu & Oliver, 2005; Nævdal et al., 2005), and hydrology (Chen & Zhang, 2006; Reichle et al., 2002). Nonlinearity and non-Gaussianity constitute the major challenges for EnKF, and many variants have been proposed for solving these complex situations. Gu and Oliver (2007) developed an iterative ensemble Kalman filter, named the ensemble randomized maximum likelihood filter (EnRML), for solving nonlinear multiphase fluid flow. Zhou et al. (2011) proposed the normal-score ensemble Kalman filter (NS-EnKF), which can handle non-Gaussian model parameters and state variables by transforming the original state vector into a new Gaussian vector.

The ensemble smoother (ES), proposed by Van Leeuwen and Evensen (1996), is another ensemble-based method for inverse modeling, which performs one global update using data from all time steps simultaneously rather than sequential updates as that in EnKF. Skjervheim and Evensen (2011) used the ES for history matching of reservoir simulation, and compared the performance of ES with EnKF. In their work, they concluded that, by avoiding the simulation restarts associated with the sequential updates in EnKF, ES is more efficient and simpler compared with EnKF, and it can provide identical results to EnKF for linear dynamic models. Performing only a single update, however, ES is improper for nonlinear dynamic problems. For solving this problem, the iterative ensemble smoother (IES) is developed to achieve better performance for nonlinear problems (Chen & Oliver, 2012, 2013). An IES method based on a modified Levenberg–Marquardt method is proposed by Chen and Oliver (2013), in which the explicit computation of the sensitivity matrix is avoided by modifying the Hessian matrix. Moreover, Chang et al. (2017) put forward a surrogate model based IES for parameter inversion, in which the polynomial chaos expansion (PCE) surrogate and the interpolation surrogate are employed to augment the efficiency of the inversion process. Similarly, Ju et al. (2018) presented an adaptive Gaussian process (GP)-based iterative ensemble smoother (GPIES) for heterogeneous conductivity estimation of subsurface flow, in which the GP surrogate is constructed and adaptively refined by adding a few new points chosen from the updated parameter ensemble.

Recently, deep-learning based approaches have been adopted for solving inverse problems, and achieved good performance in numerous fields. Jin et al. (2017) proposed the FBPConvNet for solving image reconstruction problems, which combined the Filtered Back Projection (FBP) and convolutional neural network (CNN). In their work, the FBP is first performed, and subsequently used as input for CNN to regress the FBP results to the ground truth images. The framework is adopted for X-ray computed tomography (CT) reconstruction, and good results are obtained. Adler and Öktem (2017) developed a partially learned gradient descent scheme for solving ill-posed inverse problems, in which the gradient-like iterative scheme is used for optimizing the objective function, and gradients are learned with a CNN from the training data. Antholzer et al. (2019) adopted a deep-learning framework for image reconstruction in photoacoustic tomography, in which a CNN is trained with training data and used for image reconstruction from sparse data. The CNN structure has also been utilized for seismic inversion in geophysics. Wu and Lin (2020) proposed the InversionNet for full waveform inversion, which employed an encoder-decoder structure of CNN. Li et al. (2020) proposed an end-to-end seismic inversion network (SeisInvNet), which takes advantage of all seismic data for reconstruction of velocity models. In hydrology, Mo et al. (2019) developed a deep autoregressive neural network-based surrogate as a forward groundwater contaminant transport model, and the iterative local updating ensemble smoother is adopted for groundwater contaminant source identification. Deep-learning techniques have also been used in parameterization of geological media, such as Variational Autoencoder (VAE) (Laloy et al., 2017) and Generative Adversarial Network (GAN) (Laloy et al., 2018), which constitutes an important step for geological media property inversion.

Inverse modeling based on the Physics Informed Neural Network (PINN) (Raissi et al., 2019) has also been investigated. The PINN, proposed by Raissi et al. (2019), can incorporate the residual of partial differential equations (PDEs) into the loss function of a neural network, and can be used to identify coefficients in the PDEs with available data. For space-dependent coefficients, such as hydraulic conductivity, Tartakovsky et al. (2020) used neural networks to approximate both the hydraulic conductivity and hydraulic head, which are trained with Darcy's law, in addition to available measurements. A similar strategy has also been applied to multiphysics data assimilation for subsurface transport (He et al., 2020). However, PINN-based inverse modeling methods do not honor geostatistical information of model parameters, and thus require a large number of measurements, which is unrealistic for engineering practice.

In this work, two categories of inverse modeling approaches are proposed and compared. The first category is deep-learning surrogate-based methods, including the Theory-guided neural network (TgNN) surrogate-based gradient method, iterative ensemble smoother (IES), and training method. The TgNN, developed by Wang et al. (2020b), can incorporate not only physical principles, but also practical engineering theories, and can be further developed as a surrogate for problems with uncertain model parameters (Wang et al., 2020a). Since the TgNN surrogate can be constructed with limited simulation data, and even in a label-free manner, fewer forward simulations are required. In addition, model predictions from the TgNN surrogate necessitate little computation effort, which can improve efficiency for parameter inversion. Moreover, solving inverse problems with the gradient method and the training method becomes feasible since the gradient can be easily obtained from the TgNN surrogate. The second category is the direct-deep-learning inversion method, in which TgNN with geostatistical constraint, named TgNN-geo, is proposed as the deep-learning framework for inverse modeling. In TgNN-geo, two neural networks are introduced to approximate the random model parameter and solution, respectively. In order to honor prior knowledge of geostatistical information of the random model parameter, the neural network for approximating the random model parameter is trained by using the realizations generated from Karhunen-Loeve expansion (KLE). By minimizing the loss function of TgNN-geo, estimation of the model parameter and approximation of the model solution can be simultaneously obtained. Indeed, since prior geostatistical information can be incorporated, the direct inversion method based on TgNN-geo can work well, even in cases with sparse spatial measurements. The two categories of methods are tested with several subsurface flow problems, and the advantages and disadvantages of the two categories of methods are discussed.

## 2 Inverse Problem

To introduce the inverse problem, let us consider a mathematical model with the following relationship:

$$g(\mathbf{m}) = \mathbf{d} \qquad (1)$$

where $\mathbf{m}$ denotes the model parameters; $g(\cdot)$ denotes the theoretical forward model or simulator; and $\mathbf{d}$ denotes the predicted outputs from the model, given the model parameters $\mathbf{m}$. Predicting the model response with known model parameters by running a simulator can be termed *the forward problem*. However, in many science and engineering practices, the model parameters are not specifically known, and only some measurements are available. As a consequence, one usually needs to infer the model parameters that characterize the system with given measurements to make more accurate predictions of the model response, which can be termed *the inverse problem*. The inverse problem requires solving a group of equations (Oliver

et al., 2008):

$$\mathbf{d}^{obs} = g(\mathbf{m}) + \varepsilon \tag{2}$$

where $\mathbf{d}^{obs}$ denotes the measurement data, including direct and indirect measurements; and $\varepsilon$ denotes the observation errors. The solution of the inverse problem is usually non-unique, and one may need to characterize the posterior conditional probability density function (PDF) of model parameters $\mathbf{m}$ given measurement data $\mathbf{d}^{obs}$, which is $p(\mathbf{m}|\mathbf{d}^{obs})$. To solve the inverse problem, one may maximize the posterior PDF of model parameters from a probabilistic perspective. If the prior PDF of model parameters, modeling and measurement errors all follow a Gaussian distribution, maximizing the posterior PDF is equivalent to minimizing the objective function (Oliver et al., 2008):

$$O(\mathbf{m}) = \frac{1}{2}\left(\mathbf{d}^{obs} - g(\mathbf{m})\right)^T C_D^{-1} \left(\mathbf{d}^{obs} - g(\mathbf{m})\right) \\ + \frac{1}{2}\left(\mathbf{m} - \mathbf{m}^{pr}\right)^T C_M^{-1} \left(\mathbf{m} - \mathbf{m}^{pr}\right) \tag{3}$$

where $C_D$ denotes the covariance matrix of modeling and measurement errors; $C_M$ denotes the prior covariance matrix of the model parameters; and $\mathbf{m}^{pr}$ denotes the prior estimate of the model parameters. Therefore, under certain conditions, the inverse modeling problem can be understood as an optimization problem.

## 3 Optimization Methods

### 3.1 Optimization based on gradient method

To solve the optimization problem shown in Eq. (3), the gradient method is a straightforward option. Among different types of gradient methods, the Levenberg–Marquardt algorithm has achieved good performance for optimization, and the iterative update of model parameters can be formulated as (Chang et al., 2017; Chen & Oliver, 2013):

$$\mathbf{m}_{l+1} = \mathbf{m}_l - \left[(1+\lambda_l)C_M^{-1} + G_l^T C_D^{-1} G_l\right]^{-1} \left[C_M^{-1}\left(\mathbf{m}_l - \mathbf{m}^{pr}\right) + G_l^T C_D^{-1}\left(g(\mathbf{m}_l) - \mathbf{d}^{obs}\right)\right] \tag{4}$$

where $l$ denotes the iteration number index; $G_l$ denotes the sensitivity matrix of data with respect to model parameters at $l$th iteration; and $\lambda_l$ denotes a multiplier, which modified the Hessian to reduce the influence of the large data mismatch in early iterations (Chen & Oliver, 2013). Eq. (4) can be further rewritten as (Chang et al., 2017):

$$\mathbf{m}_{l+1} = \mathbf{m}_l - \frac{1}{1+\lambda_l}\left[C_M - C_M G_l^T\left((1+\lambda_l)C_D + G_l C_M G_l^T\right)^{-1} G_l C_M\right] C_M^{-1}\left(\mathbf{m}_l - \mathbf{m}^{pr}\right) \\ - C_M G_l^T\left((1+\lambda_l)C_D + G_l C_M G_l^T\right)^{-1}\left(g(\mathbf{m}_l) - \mathbf{d}^{obs}\right) \quad (5)$$

In the iteration process of the gradient method, the sensitivity matrix of observed data with respect to model parameters should be calculated in each iteration step. For the numerical simulator, calculation of sensitivity coefficients is tedious with the adjoint method or the perturbation method. Therefore, an efficient approach for sensitivity calculation is needed. Moreover, updating one realization of model parameters is not adequate for characterizing posterior PDF.

**3.2 Optimization based on iterative ensemble smoother (IES)**

In order to characterize the posterior PDF, the ensemble-based method can be adopted. The sensitivity matrix can also be approximated with the ensemble statistics. Specifically, in the IES, the Hessian matrix is modified and approximated, in which the explicit computation of the sensitivity matrix is avoided (Chen & Oliver, 2013).

In the ensemble-based method, a group of realizations of model parameters should be updated with Eq. (5) as follows:

$$\mathbf{m}_{l+1,j} = \mathbf{m}_{l,j} - \frac{1}{1+\lambda_l}\left[C_M - C_M G_{l,j}^T\left((1+\lambda_l)C_D + G_{l,j} C_M G_{l,j}^T\right)^{-1} G_{l,j} C_M\right] C_M^{-1}\left(\mathbf{m}_{l,j} - \mathbf{m}_j^{pr}\right) \\ - C_M G_{l,j}^T\left((1+\lambda_l)C_D + G_{l,j} C_M G_{l,j}^T\right)^{-1}\left(g(\mathbf{m}_{l,j}) - \mathbf{d}_j^{obs}\right), \quad j=1,\ldots,N_e \quad (6)$$

where $j$ denotes the realization index; $G_{l,j}$ denotes the sensitivity matrix, taking a value at $\mathbf{m}_{l,j}$; and $N_e$ denotes the total number of realizations in the ensemble.

The updating formula is then further modified in the following ways: first, the prior covariance matrix $C_M$ in the Hessian matrix is replaced by the covariance matrix of the updated model parameters $C_{M_l}$ at iteration step $l$ (Chang et al., 2017; Chen & Oliver, 2013); and second, the sensitivity matrix $G_{l,j}$ is replaced by the averaged sensitivity matrix $\bar{G}_l$ at iteration step $l$ (Chang et al., 2017; Gu & Oliver, 2007; Le et al., 2016). The updating formula then becomes:

$$\mathbf{m}_{l+1,j} = \mathbf{m}_{l,j} - \frac{1}{1+\lambda_l}\left[C_{M_l} - C_{M_l}\bar{G}_l^T\left((1+\lambda_l)C_D + \bar{G}_l C_{M_l}\bar{G}_l^T\right)^{-1}\bar{G}_l C_{M_l}\right]C_M^{-1}\left(\mathbf{m}_{l,j} - \mathbf{m}_j^{pr}\right)$$
$$- C_{M_l}\bar{G}_l^T\left((1+\lambda_l)C_D + \bar{G}_l C_{M_l}\bar{G}_l^T\right)^{-1}\left(g(\mathbf{m}_{l,j}) - \mathbf{d}_j^{obs}\right), \quad j=1,\ldots,N_e \qquad (7)$$

Then, the following approximations are adopted (Chang et al., 2017; Zhang, 2001):

$$C_{M_l}\bar{G}_l^T \approx C_{M_l D_l} \qquad (8)$$

$$\bar{G}_l C_{M_l}\bar{G}_l^T \approx C_{D_l D_l} \qquad (9)$$

where $C_{M_l D_l}$ denotes the cross covariance between the updated model parameters and the predicted data at iteration step $l$; and $C_{D_l D_l}$ denotes the covariance of the predicted data at iteration step $l$. By substituting Eq. (8) and Eq. (9) into Eq. (7), one has the following:

$$\mathbf{m}_{l+1,j} = \mathbf{m}_{l,j} - \frac{1}{1+\lambda_l}\left[C_{M_l} - C_{M_l D_l}\left((1+\lambda_l)C_D + C_{D_l D_l}\right)^{-1}C_{D_l M_l}\right]C_M^{-1}\left(\mathbf{m}_{l,j} - \mathbf{m}_j^{pr}\right)$$
$$- C_{M_l D_l}\left((1+\lambda_l)C_D + C_{D_l D_l}\right)^{-1}\left(g(\mathbf{m}_{l,j}) - \mathbf{d}_j^{obs}\right), \quad j=1,\ldots,N_e \qquad (10)$$

In the IES, the uncertainty can be quantified with the updated realizations, and it is simple to combine with the simulator since explicit computation of sensitivity is not required. When performing IES, the forward simulation should be run repeatedly during the iteration because a group of realizations should be updated, which may incur large computational effort. Constructing a surrogate model for forward evaluation is an effective way to improve the efficiency of IES (Chang et al., 2017). Due to the curse of dimensionality, however, most existing surrogate models may not work well for problems with large dimensionality. In the next section, a deep-learning surrogate is introduced, which shows superiority for problems with large dimensionality.

## 4 Deep-learning Surrogate-based Inversion Methods

In this section, several deep-learning surrogate-based inversion methods will be presented. Here, we will first introduce the investigated physical problem in this work.

Consider a dynamic physical problem with the following stochastic partial differential equations (SPDEs) as the governing equation:

$$\begin{aligned} &\text{N}(h(\mathbf{x},t;\omega); K(\mathbf{x};\omega)) = 0, & \mathbf{x} \in D,\ \omega \in \Omega,\ t \in [0,T] \\ &\text{B}(h(\mathbf{x},t;\omega)) = b(\mathbf{x}), & \mathbf{x} \in \Gamma,\ t \in [0,T] \\ &\text{I}(h(\mathbf{x},0;\omega)) = i(\mathbf{x}), & \mathbf{x} \in D,\ t = 0 \end{aligned} \quad (11)$$

where N denotes a general nonlinear differential operator; $\omega$ denotes the random variables in the probability space $\Omega$; $D$ denotes the physical domain; $T$ denotes the time span; $K(\mathbf{x};\omega)$ denotes the space-dependent coefficients, which can be seen as a random field; $h(\mathbf{x},t;\omega)$ denotes the quantity of interest or the solution of the problem; B denotes the boundary condition operator defined on the boundary domain $\Gamma$; and I denotes the initial condition operator defined at initial time.

### 4.1 Parameterization of random field

As introduced previously, the space-dependent uncertain model parameter $K(\mathbf{x};\omega)$ can be regarded as a random field or random process. In order to efficiently represent and operate the model parameter, some techniques can be adopted to parameterize the random field, such as Singular Value Decomposition (SVD) (Tavakoli et al., 2011), Variational Autoencoder (VAE) (Canchumuni et al., 2019; Laloy et al., 2017), and Generative Adversarial Network (GAN) (Chan & Elsheikh, 2019; Laloy et al., 2018). In this work, Karhunen-Loeve expansion (KLE) is utilized for parameterization, which can honor second-order moments.

In this work, we assume that $\ln K(\mathbf{x};\omega)$ is a Gaussian random field. Let $Y(\mathbf{x};\omega) = \ln K(\mathbf{x};\omega)$, and using KLE, $Y(\mathbf{x};\omega)$ can be expressed as (Ghanem & Spanos, 2003; Zhang & Lu, 2004):

$$Y(\mathbf{x};\omega) = \overline{Y}(\mathbf{x}) + \sum_{i=1}^{\infty} \sqrt{\lambda_i} f_i(\mathbf{x}) \xi_i(\omega) \qquad (12)$$

where $\overline{Y}(\mathbf{x})$ denotes the mean of the random field; $\lambda_i$ and $f_i(\mathbf{x})$ denote the eigenvalue and eigenfunction of the covariance, respectively; and $\xi_i(\omega)$ denotes the independent orthogonal Gaussian random variables with zero mean and unit variance. The infinite terms in Eq. (12) can be truncated with a finite number of terms (*n*) for retaining a certain percentage of energy ($\sum_{i=1}^{n} \lambda_i / \sum_{i=1}^{\infty} \lambda_i$). The random field $Y(\mathbf{x};\omega)$ can then be expressed as follows:

$$Y(\mathbf{x};\omega) = \overline{Y}(\mathbf{x}) + \sum_{i=1}^{n} \sqrt{\lambda_i} f_i(\mathbf{x}) \xi_i(\omega) \qquad (13)$$

Therefore, the model parameter $K(\mathbf{x};\omega)$ can be parameterized by a vector composed of independent random variables:

$$\boldsymbol{\xi} = \{\xi_1(\omega), \xi_2(\omega), \cdots, \xi_n(\omega)\} \qquad (14)$$

**4.2 TgNN as deep-learning surrogate**

In this subsection, the Theory-guided Neural Network (TgNN) surrogate (Wang et al., 2020a) is constructed for the problem of Eq. (11). In the TgNN surrogate, physical/engineering constraints and other domain knowledge are incorporated as prior knowledge into the neural network training process. Considering the governing equation in Eq. (11), in order to approximate the quantity of interest $h(\mathbf{x},t;\omega)$, a Deep Neural Network (DNN) is defined as follows (with $K(\mathbf{x};\omega)$ parameterized by $\boldsymbol{\xi}$):

$$h(\mathbf{x},t;\omega) \approx \hat{h}(\mathbf{x},t,\boldsymbol{\xi};\ \theta) = NN(\mathbf{x},t,\boldsymbol{\xi};\ \theta) \qquad (15)$$

where $\theta$ denotes the parameters of the network, including weights and bias. Therefore, the location, time, and stochastic parameters comprise the inputs of the neural network, i.e., $(\mathbf{x}, t, \boldsymbol{\xi})$, as shown in **Figure 1**.

Several forward model simulations should be performed to provide training data for surrogate modeling, and the labeled data points can be represented as $\{(\mathbf{x}_i, t_i, \xi_i), h_i\}_{i=1}^{N}$, where $N$ denotes the total number of labeled training data. Then, the loss function of data mismatch between network prediction and ground truth can be formulated as mean squared error:

$$MSE_{DATA}(\theta) = \frac{1}{N}\sum_{i=1}^{N}\left|\hat{h}_i(t_i,\mathbf{x}_i,\xi_i;\ \theta) - h_i\right|^2 \qquad (16)$$

In order to achieve theory-guided training, the DNN can then be substituted into the governing equation and boundary/initial conditions in Eq. (11), and the loss functions for physics-violation can be expressed as follows:

$$MSE_{PDE}(\theta) = \frac{1}{N_c}\sum_{i=1}^{N_c}\left\{\mathrm{N}\left[\hat{h}_i(\mathbf{x}_i,t_i,\xi_i;\ \theta); \overline{Y}(\mathbf{x}) + \sum_{i=1}^{n}\sqrt{\lambda_i}f_i(\mathbf{x})\xi_i\right]\right\}^2 \qquad (17)$$

$$MSE_B(\theta) = \frac{1}{N_B}\sum_{i=1}^{N_B}\left|\mathrm{B}(\hat{h}_i(\mathbf{x}_i^b,t_i^b,\xi_i^b;\ \theta)) - b(\mathbf{x}_i^b)\right|^2 \qquad (18)$$

$$MSE_I(\theta) = \frac{1}{N_I}\sum_{i=1}^{N_I}\left|\mathrm{I}(\hat{h}_i(\mathbf{x}_i^0,0,\xi_i^0;\ \theta)) - i(\mathbf{x}_i^0)\right|^2 \qquad (19)$$

where $\{\mathbf{x}_i, t_i, \xi_i\}_{i=1}^{N_c}$ denote the collocation points to enforce the physical constraints (Raissi et al., 2019); $\{\mathbf{x}_i^b, t_i^b, \xi_i^b\}_{i=1}^{N_B}$ denote the collocation points to enforce the boundary conditions; $\{\mathbf{x}_i^0, 0, \xi_i^0\}_{i=1}^{N_I}$ denote the collocation points at the initial time to enforce the initial conditions; and $N_c$, $N_B$, and $N_I$ denote the total number of collocation points for governing equation, boundary conditions, and initial conditions, respectively.

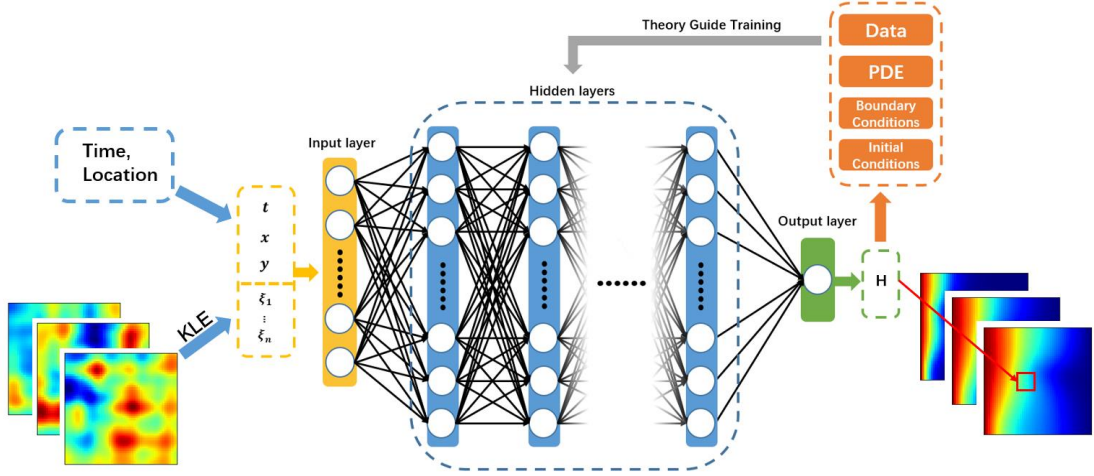

**Figure 1.** Structure of the TgNN surrogate.

Consequently, in order to train the TgNN surrogate, those loss functions can be optimized simultaneously by minimizing a total loss function, which is defined as:

$$L(\theta) = \lambda_{DATA} MSE_{DATA}(\theta) + \lambda_{PDE} MSE_{PDE}(\theta) \\ + \lambda_{B} MSE_{B}(\theta) + \lambda_{I} MSE_{I}(\theta) \tag{20}$$

where $\lambda_{DATA}$, $\lambda_{PDE}$, $\lambda_{B}$, and $\lambda_{I}$ control the importance of each term in the total loss function. The optimization can be performed via various algorithms, such as Stochastic Gradient Descent (SGD) (Bottou, 2010), Adaptive Moment Estimation (Adam) (Kingma & Ba, 2015), etc. The framework of the TgNN surrogate is presented in **Figure 1**. By incorporating physical laws, the TgNN surrogate can be constructed with limited labeled data, or even in a label-free manner. Furthermore, the effect of the number of labeled data and collocation points has been studied in Wang et al. (2020a).

**4.3 TgNN surrogate-based inversion methods**

Using the TgNN surrogate, the model predictions for different model parameter realizations can be easily obtained. Therefore, the inversion tasks may be performed efficiently using the TgNN surrogate. In this work, three different ways to combine the inversion method with the TgNN surrogate are proposed, which will be discussed in the following.

*4.3.1 TgNN surrogate-based gradient method*

In this work, by using KLE, the independent random variables $\xi$ constitute the model parameter **m**. In addition, forward model prediction can be obtained from the constructed TgNN surrogate, i.e., $g^{\text{surr}}(\xi_l)$. Therefore, the gradient method for parameter inversion with Eq. (5) can be rewritten as follows:

$$\xi_{l+1} = \xi_l - \frac{1}{1+\lambda_l}\left[C_\xi - C_\xi G_l^T\left((1+\lambda_l)C_D + G_l C_\xi G_l^T\right)^{-1} G_l C_\xi\right]C_\xi^{-1}(\xi_l - \xi^{pr})$$
$$- C_\xi G_l^T\left((1+\lambda_l)C_D + G_l C_\xi G_l^T\right)^{-1}\left(g^{\text{surr}}(\xi_l) - \mathbf{d}^{obs}\right) \quad (21)$$

Since the TgNN surrogate is constructed with the neural network, the sensitivity matrix $G_l$ can be calculated through automatic differentiation, which is more efficient than the adjoint method or the perturbation method for sensitivity calculation. Using the gradient method, we may only update one realization of model parameter, and thus this method may have high efficiency. However, the posterior PDF cannot be characterized with only one updated realization. On the other hand, the performance of the gradient method is easily influenced by the initial guess of model parameters. For solving these problems, a group of realizations can be updated with Eq. (21), and the sensitivity matrix $G_l$ may be replaced by the average sensitivity $\bar{G}_l$. Then, Eq. (21) can be rewritten as:

$$\xi_{l+1,j} = \xi_{l,j} - \frac{1}{1+\lambda_l}\left[C_\xi - C_\xi \bar{G}_l^T\left((1+\lambda_l)C_D + \bar{G}_l C_\xi \bar{G}_l^T\right)^{-1} \bar{G}_l C_\xi\right]C_\xi^{-1}(\xi_{l,j} - \xi^{pr})$$
$$- C_\xi \bar{G}_l^T\left((1+\lambda_l)C_D + \bar{G}_l C_\xi \bar{G}_l^T\right)^{-1}\left(g^{\text{surr}}(\xi_{l,j}) - \mathbf{d}_j^{obs}\right), \quad j=1,\ldots,N_e \quad (22)$$

The uncertainty of the estimated posterior can then be quantified with updated realizations, and the stability of the algorithm can be improved.

During the iteration, the data mismatch between the measurements and predicted data from the TgNN surrogate is utilized to assess the accuracy of the updated model parameters, which is defined as follows:

$$MIS(\xi) = \frac{1}{N_e}\frac{1}{N_d}\sum_{j=1}^{N_e}\sum_{i=1}^{N_d}\left|\left(g^{\text{surr}}(\xi_j)\right)_i - \mathbf{d}_{j,i}^{obs}\right|^2 \quad (23)$$

where $N_d$ denotes the total number of measurement data. The terminating criteria for iteration are defined as:

(1) $MAX_{1 \leq j \leq Ne; 1 \leq i \leq n} \left| \xi_{l+1,j}^i - \xi_{l,j}^i \right| < \varepsilon_1$;

(2) $\dfrac{\left| MIS(\xi_{l+1}) - MIS(\xi_l) \right|}{\max(1, \left| MIS(\xi_l) \right|)} < \varepsilon_2$;

(3) Iteration reaches the pre-given maximum iteration number $I_{\max}$.

where $\varepsilon_1$ and $\varepsilon_2$ are the predefined limits of error; and $n$ denotes the dimension of the model parameter.

### 4.3.2 TgNN surrogate-based IES

Using the TgNN surrogate, Eq. (10) of IES can be reformulated as:

$$\xi_{l+1,j} = \xi_{l,j} - \frac{1}{1+\lambda_l}\left[ C_{\xi_l} - C_{\xi_l D_l}^{\text{surr}} \left( (1+\lambda_l) C_D + C_{D_l D_l}^{\text{surr}} \right)^{-1} C_{D_l \xi_l}^{\text{surr}} \right] C_\xi^{-1} \left( \xi_{l,j} - \xi_j^{pr} \right) \\ - C_{\xi_l D_l}^{\text{surr}} \left( (1+\lambda_l) C_D + C_{D_l D_l}^{\text{surr}} \right)^{-1} \left( g^{\text{surr}}(\xi_{l,j}) - \mathbf{d}_j^{\text{obs}} \right), \quad j = 1, \ldots, N_e \quad (24)$$

where $C_{\xi_l D_l}^{\text{surr}}$ and $C_{D_l D_l}^{\text{surr}}$ can be approximated with:

$$C_{\xi_l D_l}^{\text{surr}} = \frac{1}{N_e - 1} \sum_{j=1}^{N_e} \left\{ \left[ \xi_{l,j} - <\xi_{l,j}> \right] \times \left[ g^{\text{surr}}(\xi_{l,j}) - <g^{\text{surr}}(\xi_{l,j})> \right]^T \right\} \quad (25)$$

$$C_{D_l D_l}^{\text{surr}} = \frac{1}{N_e - 1} \sum_{j=1}^{N_e} \left\{ \left[ g^{\text{surr}}(\xi_{l,j}) - <g^{\text{surr}}(\xi_{l,j})> \right] \\ \times \left[ g^{\text{surr}}(\xi_{l,j}) - <g^{\text{surr}}(\xi_{l,j})> \right]^T \right\} \quad (26)$$

Since the independent random variables in $\xi$ are Gaussian random variables with zero mean and unit variance, the initial covariance $C_{\xi_0}$ in the Hessian matrix and the $C_\xi$ out of the Hessian matrix should be identity matrixes. Therefore, $C_\xi^{-1}$ can be neglected in Eq. (24). Moreover, the termination criteria for iteration can be defined in a similar way, as shown in section 4.3.1. In the IES, since a group of realizations are updated simultaneously, the influence of initial guesses is weak. The explicit computation of sensitivity is avoided by using the covariance. Since the covariance can be calculated from the ensemble, the implementation of IES is simple. Some errors, however, will be incurred from the approximations shown in Eq. (8) and Eq. (9). Using the TgNN surrogate, the computational cost of forward model prediction during the iteration can be significantly reduced, and the efficiency of the inversion process can be improved.

### 4.3.3 TgNN surrogate-based training method

The above inversion methods are designed by following the optimization framework, which has similar procedures to that in deep-learning. In the deep-learning framework, a loss function is designed and minimized during the training process. Here, following the deep-learning framework, another inversion method is proposed, i.e., the training method. With the constructed TgNN surrogate $\hat{h}(\mathbf{x},t,\xi;\ \theta)$, the objective function Eq. (3) for inverse modeling can be optimized directly as follows:

$$O(\xi) = \frac{1}{2}\left(\mathbf{d}^{obs} - g^{\text{surr}}(\xi)\right)^T C_D^{-1} \left(\mathbf{d}^{obs} - g^{\text{surr}}(\xi)\right) \\ + \frac{1}{2}\left(\xi - \xi^{pr}\right)^T C_M^{-1} \left(\xi - \xi^{pr}\right) \tag{27}$$

where $g^{\text{surr}}(\xi)$ denotes the prediction from the TgNN surrogate. While in the training process, the parameters $\theta$ of the neural network are fixed, including weights and bias, and the model parameters $\xi$, i.e., the input of the TgNN surrogate, are updated to optimize the objective function. Various optimization algorithms can be utilized to minimize the objective function, and the algorithm Adam (Kingma & Ba, 2015) is adopted here, which can be implemented easily with existing deep-learning frameworks, such as Pytorch (Paszke et al., 2017) and tensorflow (Abadi et al., 2016). This procedure is similar to the neural network training process, and the difference is that the tunable parameters are the inputted random variables, and not the weights of the network. Compared with the gradient method or the IES method introduced previously, neither the Gauss-Newton nor the Levenberg–Marquardt algorithm is adopted to optimize the objective function, and some built-in algorithms of Pytorch are utilized instead. Therefore, the parameter inversion tasks can be simply implemented with many optimization algorithms embedded in the existing deep-learning frameworks, and one only needs to set some parameters, such as the training iterations and the learning rate. It is worth noting that the training method can be applied benefiting from the constructed TgNN surrogate, which is derivable and efficient to evaluate.

## 5 Direct-deep-learning-inversion Methods

In this section, inversion methods directly via the deep-learning framework are introduced. Different from the deep-learning surrogate-based inversion methods, in this category of methods, it is not necessary to construct the surrogate, and both the measurements of SPDE solutions $h(\mathbf{x},t;\omega)$ and the measurements of model parameters $K(\mathbf{x};\omega)$ are usually required.

### 5.1 Direct-deep-learning-inversion method without incorporating geostatistics

Inversion methods directly via deep-learning framework have previously been investigated. Raissi et al. (2019) recently proposed the Physics Informed Neural Network (PINN) framework, which is used to infer constant model parameters in the PDEs. In their work, a residual of the governing equation is incorporated into the loss function, and the neural network for approximating the solution, as well as the model parameters, are learned together during the training process. Tartakovsky et al. (2020) improved the PINN framework to infer heterogeneous parameters, in which both the model parameter field and the SPDE solution are approximated with neural networks and substituted into the governing equation to constitute the physics-constrained loss. In this subsection, we will briefly introduce the direct PINN inversion method, additional details of which can be found in Raissi et al. (2019) and Tartakovsky et al. (2020).

Consider the problem with governing equation Eq. (11), assuming $N_k$ measurements of the model parameter field and $N_h$ measurements of the SPDE solution are collected, and two fully-connected feed forward neural networks are defined to approximate the model parameter and solution, respectively:

$$\hat{K}(\mathbf{x};\ \theta_k) = NN_k(\mathbf{x};\ \theta_k) \tag{28}$$

$$\hat{h}(\mathbf{x},t;\ \theta_h) = NN_h(\mathbf{x},t;\ \theta_h) \tag{29}$$

Then, the two neural networks can be trained simultaneously by minimizing the loss function:

$$\begin{aligned}L(\theta_k,\theta_h) = &\frac{1}{N_f}\sum_{i=1}^{N_f}\left\{\mathrm{N}\left[\hat{h}_i(\mathbf{x}_i,t_i;\ \theta_h); \hat{K}_i(\mathbf{x}_i;\ \theta_k)\right]\right\}^2 \\&+ \frac{1}{N_b}\sum_{i=1}^{N_b}\left|\mathrm{B}(\hat{h}_i(\mathbf{x}_i^b,t_i^b;\ \theta_h)) - b(\mathbf{x}_i^b)\right|^2 \\&+ \frac{1}{N_i}\sum_{i=1}^{N_i}\left|\mathrm{I}\ (\hat{h}_i(\mathbf{x}_i^0,0;\ \theta_h)) - i(\mathbf{x}_i^0)\right|^2 \\&+ \frac{1}{N_k}\sum_{i=1}^{N_k}\left|\hat{K}_i(\mathbf{x}_i;\ \theta_k) - K_i\right|^2 + \frac{1}{N_h}\sum_{i=1}^{N_h}\left|\hat{h}_i(\mathbf{x}_i,t_i;\ \theta_h) - h_i\right|^2\end{aligned} \tag{30}$$

where $N_f$, $N_b$, and $N_i$ denote the number of collocation points for PDE, boundary conditions, and initial conditions, respectively. The first three terms in the loss function $L(\theta_k,\theta_h)$ constitute the physics-based loss, and the last two terms represent the data-driven loss or the data mismatch (Tartakovsky et al., 2020).

Although the loss function is similar to that for TgNN surrogate construction, i.e., Eq. (20), the following two differences exist: first, the PINN model is constructed for a specific conductivity field, and thus no random variables exist in the neural network inputs; and second, two different neural networks are defined in the PINN model with two sets of network parameters ($\theta_k,\theta_h$) needing to be trained simultaneously. The two neural networks can be trained with optimization algorithms, such as Adam (Kingma & Ba, 2015). Once trained, the network $\hat{K}(\mathbf{x};\ \theta_k) = NN_k(\mathbf{x};\ \theta_k)$ can provide the estimation of the model parameter field, and $\hat{h}(\mathbf{x},t;\ \theta_h) = NN_h(\mathbf{x},t;\ \theta_h)$ can approximate the solution.

The PINN method has attracted much attention for forward modeling, and has also been used for multiphysics data assimilation in subsurface transport (He et al., 2020). However, two drawbacks exist for this method: first, geostatistical information cannot be honored when approximating the model parameter field with a neural network; and second, a relatively large number of direct measurements of model parameter are required to obtain satisfactory accuracy, which is not realistic in engineering practice. It is worth noting that the TgNN surrogate-based methods described in subsection 4.3 work well even without direct model parameter measurements.

**5.2 Direct-deep-learning-inversion method constrained with geostatistics**

For solving the problems with PINN, in this subsection, a direct TgNN inversion method is proposed, which can incorporate geostatistical information of the model parameter field and can reduce dependence on the number of direct measurement data. The proposed method utilizes physical/engineering principles and domain knowledge, as well as geostatistical information, and it is named TgNN-geo in this work. For differentiation, the PINN described in section 5.1 is named PINN-no-geo in this work.

Regarding geostatistical information, KLE provides an effective method for honoring second-order moments. Therefore, in order to incorporate geostatistical information in TgNN, a fully-connected neural network can be defined to approximate the relationship between the random variables $\xi$ and the random parameter field $K(\mathbf{x})$, with the realizations generated from KLE serving as training data. Here, it is necessary to mention that, if only some training images or realizations are available serving as geostatistical information without knowing the theoretical geostatistical model, certain techniques, such as Variational Autoencoder (VAE) (Laloy et al., 2017) and Generative Adversarial Network (GAN) (Laloy et al., 2018), can be adopted to construct a mapping from training images/realizations $K(\mathbf{x})$ to the latent variables $\xi$. For approximating random model parameter fields, the neural network can then be defined as:

$$\hat{K}(\mathbf{x},\xi;\ \theta_{para}) = NN_{para}(\mathbf{x},\xi;\ \theta_{para}) \tag{31}$$

Therefore, the loss function can be expressed as:

$$MSE_{para}(\theta_{para}) = \frac{1}{N_{para}} \sum_{i=1}^{N_{para}} \left| \hat{K}(\mathbf{x},\xi;\ \theta_{para}) - K(\mathbf{x}_i,\xi_i) \right|^2 \tag{32}$$

where $K(\mathbf{x}_i, \xi_i)$ denotes data points sampled in the generated realizations of model parameters from KLE; and $N_{para}$ denotes the total number of sampled data points. Subsequently, after the minimization of Eq. (32), the trained neural network $NN_{para}(\mathbf{x}, \xi; \theta_{para})$ can learn the information of spatial correlation from the generated KLE realizations.

The loss function of the direct TgNN inversion method can be written as follows:

$$L(\xi_k, \theta_h) = \frac{1}{N_f} \sum_{i=1}^{N_f} \left\{ \mathrm{N}\left[\hat{h}_i(\mathbf{x}_i, t_i; \theta_h); \hat{K}_i(\mathbf{x}_i; \xi_k)\right]\right\}^2$$
$$+ \frac{1}{N_b} \sum_{i=1}^{N_b} \left| \mathrm{B}(\hat{h}_i(\mathbf{x}_i^b, t_i^b; \theta_h)) - b(\mathbf{x}_i^b) \right|^2$$
$$+ \frac{1}{N_i} \sum_{i=1}^{N_i} \left| \mathrm{I}(\hat{h}_i(\mathbf{x}_i^0, 0; \theta_h)) - i(\mathbf{x}_i^0) \right|^2 \quad (33)$$
$$+ \frac{1}{N_k} \sum_{i=1}^{N_k} \left| \hat{K}_i(\mathbf{x}_i; \xi_k) - K_i \right|^2 + \frac{1}{N_h} \sum_{i=1}^{N_h} \left| \hat{h}_i(\mathbf{x}_i, t_i; \theta_h) - h_i \right|^2$$

It is worth noting that while training the loss function $L(\xi_k, \theta_h)$, the parameters $\theta_{para}$ of $NN_{para}(\mathbf{x}, \xi; \theta_{para})$ are fixed, and the input random variables $\xi_k$ are tunable. After finishing the training process, the estimated model parameter field can be obtained from $NN_{para}(\mathbf{x}, \xi; \theta_{para})$ by inputting the trained $\xi_k$. There are two reasons that we use a neural network to approximate the random model parameter in Eq. (30) rather than using KLE directly: first, the differential operator can be easily implemented working with a neural network; and second, analytical expressions of KLE are not always available for given correlation structures.

Compared with PINN-no-geo, in TgNN-geo, the network for approximating the model parameter is first trained by the realizations generated with KLE or other random field generators, which contain spatial correlation information. Moreover, during the inversion process (training the two networks simultaneously), only the inputted random variables $\xi_k$ need to be updated for the $NN_{para}(\mathbf{x}, \xi; \theta_{para})$ network, and thus the parameters of networks needed to be trained are significantly reduced.

# 6 Cases Studies

Although the proposed deep-learning based inverse modeling methods are applicable to a broad range of problems, we would like to illustrate the performance of these inversion methods with subsurface flow problems. In this section, several subsurface flow examples are designed to test the performance of the proposed two categories of deep-learning based inversion methods.

## 6.1 Subsurface flow problem

In this work, transient saturated subsurface flow problems are considered, which have a general form of governing equation:

$$S_s \frac{\partial h(\mathbf{x},t)}{\partial t} - \nabla \cdot (K(\mathbf{x})\nabla h(\mathbf{x},t)) = 0 \tag{34}$$

where $S_s$ denotes the specific storage; $h(\mathbf{x},t)$ denotes the hydraulic head; and $K(\mathbf{x})$ denotes the hydraulic conductivity. Due to the spatial variation and limited direct measurements of $K(\mathbf{x})$, large uncertainty usually exists about $K(\mathbf{x})$, and thus in the solution $h(\mathbf{x},t)$. As the model parameters can be treated as random variables/fields, the governing equation becomes a stochastic partial differential equation (SPDE). Furthermore, inverse modeling using direct and indirect measurements is necessary for accurately inferring $K(\mathbf{x})$ and predicting $h(\mathbf{x},t)$.

## 6.2 TgNN surrogate-based IES

Consider a two-dimensional dynamic subsurface flow problem, which satisfies the governing equation of Eq. (34), and is subjected to the following boundary and initial conditions:

$$h|_{x=x_0} = 202[L], h|_{x=x_0+L_x} = 200[L] \tag{35}$$

$$\left.\frac{\partial h}{\partial y}\right|_{y=y_0 \text{ or } y_0+L_y} = 0 \tag{36}$$

$$h|_{t=0, x \neq x_0} = 200[L] \tag{37}$$

where $(x_0, y_0)$ denotes the starting position of the domain; and $L_x$ and $L_y$ denote the length of the domain in the respective directions. The domain is a square, and the length in both directions takes a value of $L_x = L_y = 1020\,[L]$ (where $[L]$ denotes any consistent length unit). The specific storage is assumed to be a constant, taking a value of $S_s = 0.0001\,[L^{-1}]$. The log hydraulic conductivity $\ln K$ is assumed to be a stationary Gaussian random field with the mean and variance given as $\langle \ln K \rangle = 0$ and $\sigma_{\ln K}^2 = 1.0$, respectively. A separable exponential covariance function is defined for $\ln K$:

$$C_{\ln K}(\mathbf{x}_1, \mathbf{x}_2) = \sigma_{\ln K}^2 \exp\left\{-\left[\left(\frac{|x_1 - x_2|}{\eta_x}\right)^2 + \left(\frac{|y_1 - y_2|}{\eta_y}\right)^2\right]\right\} \tag{38}$$

where $\mathbf{x}_1 = (x_1, y_1)$ denotes the coordinate of a point in the domain. This particular covariance function is used, which admits an analytical expression for the KLE (Zhang & Lu, 2004). However, the framework proposed is applicable to any covariance function, although numerical evaluation of the KLE may be needed. The correlation length in both directions is set as $\eta_x = \eta_y = 0.4 L_x = 408\,[L]$. The $\ln K$ field is parameterized via KLE, and 20 terms are retained in the expansion, resulting in 80% energy maintained. The $\ln K$ field can then be represented by 20 standard Gaussian random variables, $\boldsymbol{\xi} = \{\xi_1(\omega), \xi_2(\omega), \cdots, \xi_{20}(\omega)\}$. A randomly generated $\ln K$ field is chosen as the reference field for the inversion target, as shown in **Figure 2(a)**.

With the given reference $\ln K$ field, the flow problem is solved by using MODFLOW software. For numerical simulation, the domain is uniformly discretized into 51×51 grids, and the total time span is chosen as $10\,[T]$ (where $[T]$ denotes any consistent time unit), and is evenly discretized into 50 time-steps with each time step being $0.2\,[T]$. Five observation points are set in the domain, as shown in **Figure 2(b)**, where the hydraulic head measurements are continuously collected for the 50 time-steps. Noises are added to the simulation data as observation errors. The noises are assumed to be uncorrelated, and thus the covariance of observation errors constitutes a diagonal matrix. The mean and standard deviation of noises are set to be 0 and 0.01, respectively.

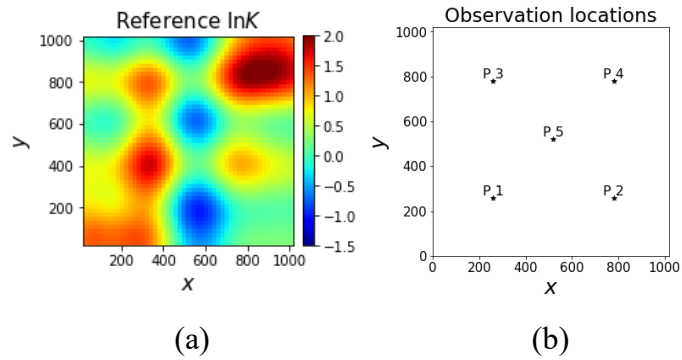

(a)            (b)

**Figure 2.** Reference field (a) and location of observation points (b).

### *6.2.1 TgNN surrogate construction*

The TgNN surrogate is first constructed. A seven-hidden-layer fully-connected neural network with 50 neurons per layer is chosen for TgNN surrogate construction. 30 hydraulic conductivity fields are generated with KLE, and numerical simulations are then performed to obtain training data. Furthermore, $10^6$ collocation points, at which the physical constraints are

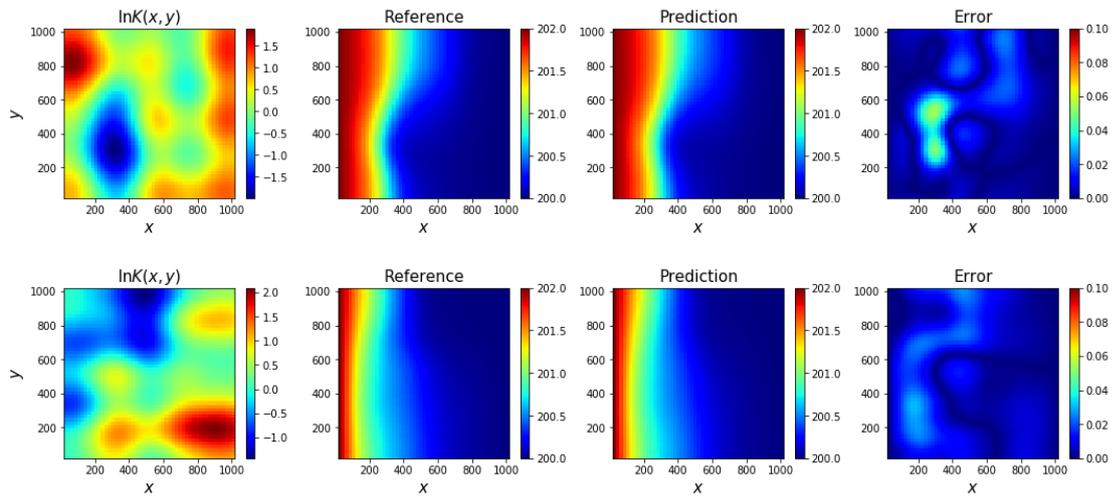

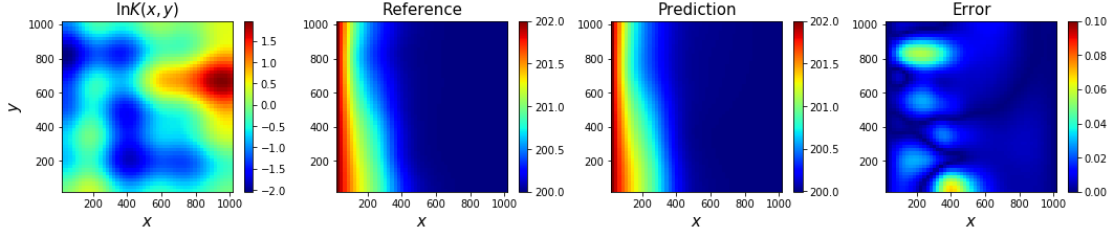

**Figure 3.** Reference and prediction from the TgNN surrogate at time-step 25 for three different hydraulic conductivity fields.

imposed, are randomly sampled from high-dimensional parameter space:

$$\begin{aligned} t_i &\sim U(t_0, t_{end}) \\ x_i &\sim U(x_0, x_0 + L_x) \\ y_i &\sim U(y_0, y_0 + L_y) \\ \xi_{1i} &\sim N(0,1) \\ &\vdots \\ \xi_{20i} &\sim N(0,1) \end{aligned} \tag{39}$$

where $U$ and $N$ denote uniform distribution and normal distribution, respectively. The Adam algorithm (Kingma & Ba, 2015) is utilized to train the neural network with a constant learning rate of 0.001 for 2,000 epochs. The training process is performed on an NVIDIA GeForce RTX 2080 Ti Graphics Processing Unit (GPU) card, which takes approximately 2.2 h (7,857.762 s). The constructed TgNN surrogate can predict the hydraulic head distribution for different conductivity fields with a relatively high accuracy, three of which are shown in **Figure 3**.

### 6.2.2 Inversion results with TgNN surrogate-based IES

The TgNN surrogate-based IES can then be implemented for inverse modeling. We set $N_e = 100$, $\varepsilon_1 = 0.01$, $\varepsilon_2 = 0.0001$, and $I_{max} = 10$. The update terminates at the fifth iteration. The mean of the initial, as well as the final updated $\ln K$ realizations are presented in **Figure 4**. It can be seen that the estimated $\ln K$ is similar to the reference field, and the major features of the reference field have been captured. In order to analyze the uncertainty of the posterior $\ln K$, the standard deviation of the ensemble at the initial and the final step are also shown in **Figure 4**. It can be seen that the standard deviation of the initial ensemble is relatively large since the realizations are randomly generated with KLE. In addition, the standard deviation largely decreases to a low level at the final updating step, which means that the final estimation has low uncertainty. It is worth noting that, since the areas near the boundaries are far from the observation points, higher uncertainty exists in these regions.

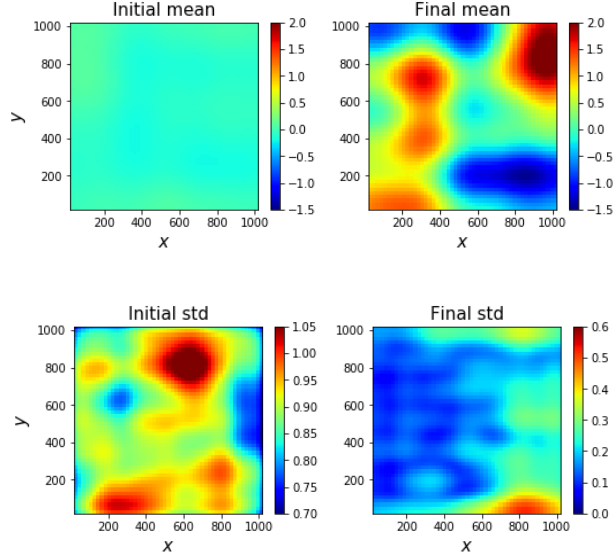

**Figure 4.** Mean and standard deviation of log hydraulic conductivity at initial and final step.

### 6.2.3 Effect of ensemble size

The sensitivity of the proposed algorithm for ensemble size is investigated in this subsection. TgNN surrogate-based IES is implemented with different ensemble sizes using the previous case. In order to evaluate the accuracy of the estimation quantitatively, the *RMSE* (root mean square error) is introduced as a criterion:

$$RMSE = \sqrt{\frac{1}{N_{cell}} \sum_{i=1}^{N_{cell}} (\ln K^{ref} - \ln K^{est})^2} \qquad (40)$$

where the superscripts *ref* and *est* denote the reference and estimation, respectively; and $N_{cell}$ denotes the total number of grid blocks. The *RMSE* of the estimated $\ln K$ from the TgNN surrogate-based IES with different $N_e$ and the corresponding standard deviation for 50 different parameter initializations are shown in **Figure 5(a)**. It can be seen that *RMSE* and the corresponding standard deviation decrease as the size of the ensemble increases, which means that the results become increasingly accurate and stable as the ensemble size gets larger.

This can be explained by the fact that the approximation of covariance becomes more accurate as the ensemble size increases.

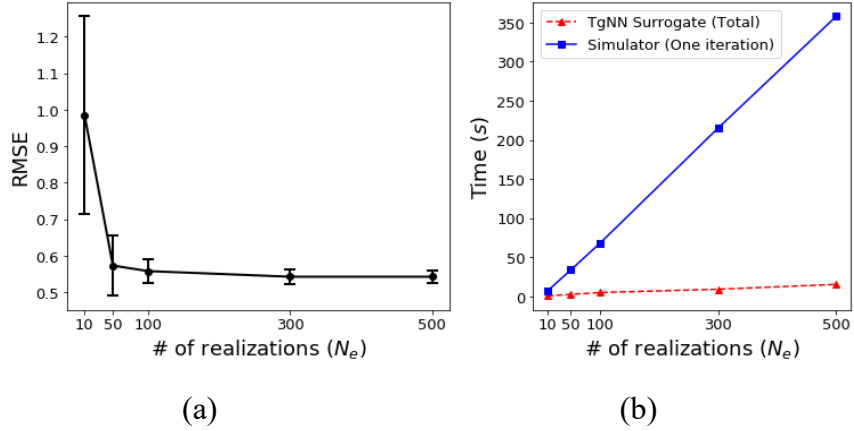

(a)          (b)

**Figure 5.** Results for ensemble size effect: (a) mean and standard deviation of RMSE for different numbers of realizations; (b) computation time for different numbers of realizations with the TgNN surrogate and simulator.

To explore the efficiency of the TgNN surrogate-based IES, the inversion consumed time for different ensemble sizes is shown in **Figure 5(b)**. For comparison, the computational cost with MODFLOW for different ensemble sizes is also provided. It can be seen that the computation time increases as the size of the ensemble increases. However, the elapsed time for inversion with the TgNN surrogate-based IES increases more slowly compared with running the simulator directly, even for just one iteration, which demonstrates the efficiency of the proposed algorithm. Indeed, the TgNN surrogate construction is time-consuming, but once trained, it can be used for parameter inversion of different cases.

**6.3 TgNN surrogate-based gradient method**

Consider the case introduced in subsection 6.2, and the TgNN surrogate-based gradient method is implemented. In this method, both the forward evaluation and sensitivity coefficients calculation are implemented with the TgNN surrogate. We set $\varepsilon_1=0.01$, $\varepsilon_2=0.0001$, and $I_{\max}=10$. The update process terminates at the third iteration, taking only 0.1234 s. The initial and final updated hydraulic conductivity field are shown in **Figure 6**. It can be seen that the final updated $\ln K$ exhibits a similar pattern to the reference field with $RMSE=0.628$, which shows that the estimation achieves satisfactory accuracy. However, unlike the ensemble-based method, only one realization is updated in the gradient method, and thus uncertainty quantification of the posterior model parameter cannot be performed. Additional discussion of this issue will be provided in the next subsection.

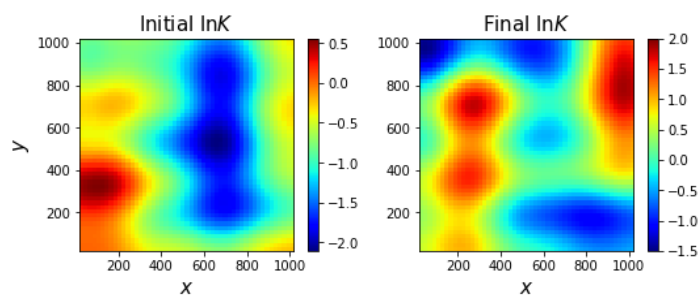

**Figure 6.** Initial and final log hydraulic conductivity from the gradient method.

### *6.3.1 Effect of initialization*

The stability of the gradient method regarding different initializations of model parameters is investigated in this subsection. Inversion tasks are performed with different initial guesses of hydraulic conductivity, the results of which are presented in **Figure 7** and **Table 1**. It can be seen that the gradient method exhibits relatively poor stability, and performance is affected by the initialization of the hydraulic conductivity. For solving this problem, a strategy is adopted, as introduced in subsection 4.3.1, which is similar to ensemble-based method. A group of realizations ($N_e = 100$) are updated with the gradient method, and the average sensitivity matrix $\bar{G}_l$ is utilized when updating the model parameters. Five groups of realizations with different initial guesses are updated, and the results are shown in **Table 2**. It can be seen that the *RMSE* for different groups are relatively stable. The mean and standard deviation of the initial and final updated realizations in Group 1 are presented in **Figure 8**, from which one can see that the uncertainty of posterior can be quantified with this method. The consumed time for the inversion process, however, increases significantly compared with updating only one realization. Moreover, the mean and standard deviation of *RMSE* for 50 different group initializations with different realization numbers are shown in **Figure 9**. It can be seen that the results become increasingly accurate and stable as the number of realizations increases, which is similar to the results of the TgNN surrogate-based IES.

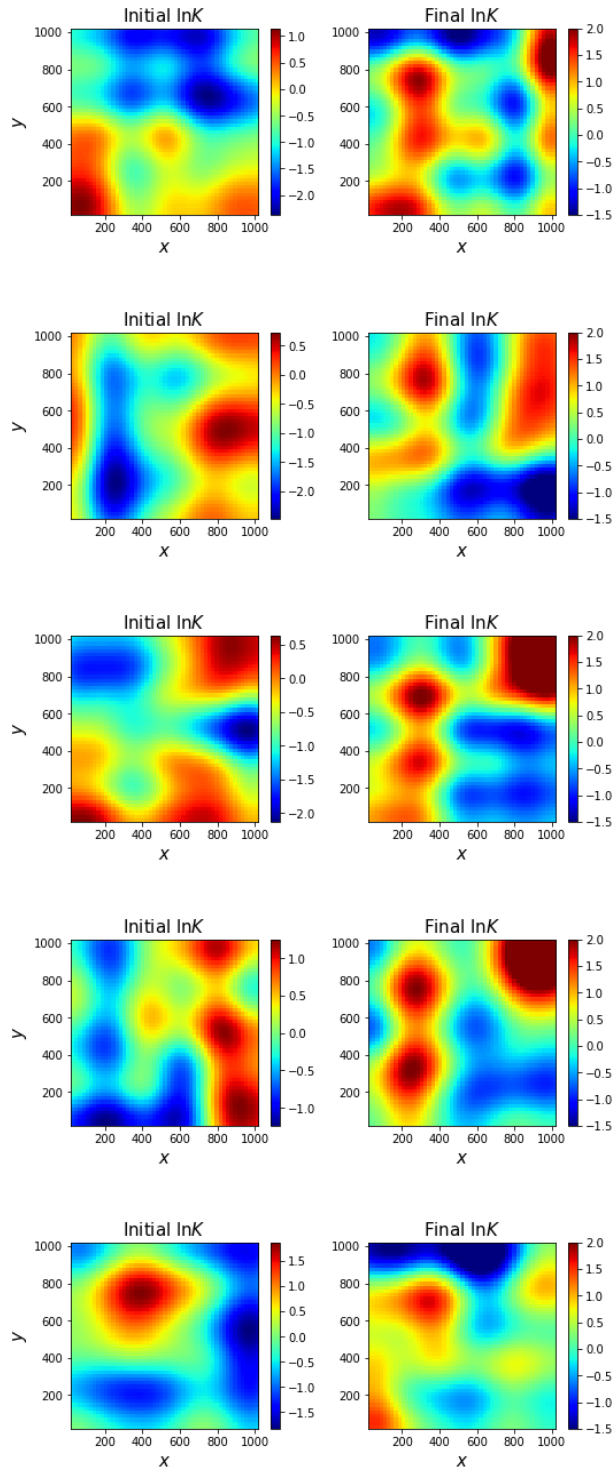

**Figure 7.** Initial and final updated log hydraulic conductivity with different parameter initializations.

**Table 1.** Inversion results for different initializations of model parameter.

|  | RMSE | Iteration | Time (s) |
|---|---|---|---|
| Initialization 1 | 0.710 | 4 | 0.1185 |
| Initialization 2 | 0.758 | 7 | 0.2108 |
| Initialization 3 | 0.699 | 4 | 0.1427 |
| Initialization 4 | 0.625 | 5 | 0.1523 |
| Initialization 5 | 0.892 | 3 | 0.1075 |

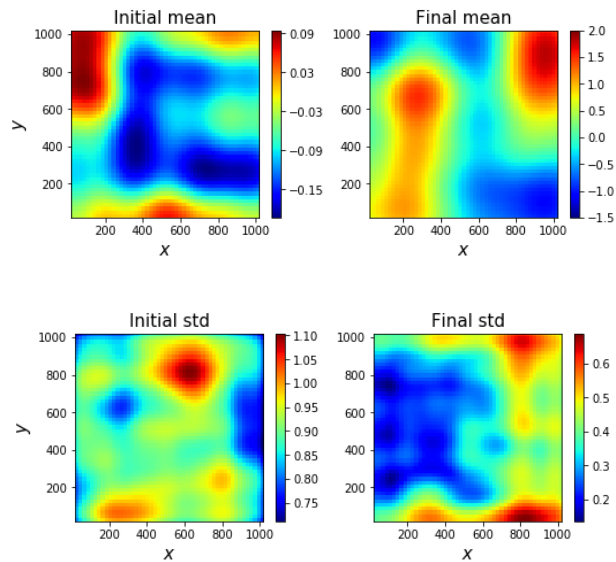

**Figure 8.** Mean and standard deviation of the initial and final updated realizations in Group 1.

**Table 2.** Inversion results for different group initializations of model parameter.

| Initialization | Realization Number | RMSE | Iteration | Time (s) |
|---|---|---|---|---|
| Group Initialization 1 | 100 | 0.575 | 3 | 5.5757 |
| Group Initialization 2 | 100 | 0.567 | 3 | 5.5680 |
| Group Initialization 3 | 100 | 0.569 | 3 | 5.5725 |
| Group Initialization 4 | 100 | 0.562 | 3 | 5.4860 |
| Group Initialization 5 | 100 | 0.558 | 3 | 5.6855 |

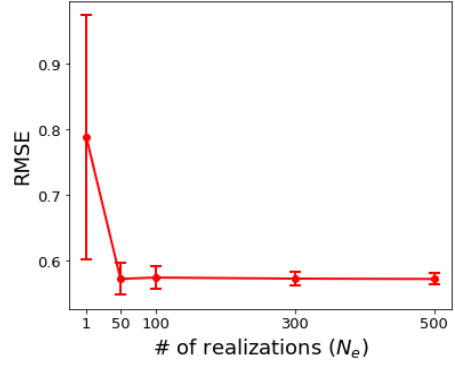

**Figure 9.** Mean and standard deviation of RMSE for 50 different group initializations with different numbers of realizations.

### 6.4 TgNN surrogate-based training method

In this subsection, parameter inversion is performed with the TgNN surrogate-based training method for the same case discussed in subsections 6.2 and 6.3. The $\ln K$ field is initialized with the mean field, as shown in **Figure 10(a)**. The model parameters $\xi$ are tuned in the training process. The Adam algorithm is adopted to minimize the objective function with learning rate of 0.1 for 300 iterations. Similar to the TgNN surrogate-based gradient method in subsection 6.3, only one realization is updated to fit the training data. The *RMSE* of estimation is 0.662, and the training process only takes 7.0109 s. The estimated $\ln K$ field is shown in **Figure 10(b)**, which demonstrates the satisfactory performance of this method. In addition, the change of objective function value during the training process is shown in **Figure 10(c)**. It is worth noting that this method can also be implemented in an ensemble manner, as introduced in subsection 6.3.1, to quantify the posterior uncertainty and stabilize the method. This case validates the feasibility of the proposed TgNN surrogate-based training method.

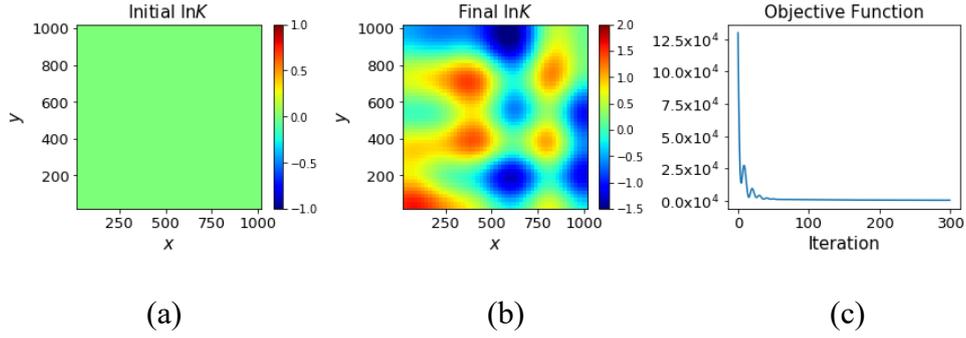

(a)                                      (b)                                  (c)

**Figure 10.** The initial log hydraulic conductivity (a), the final log hydraulic conductivity with training method (b), and the objective function in the training process (c).

## 6.5 Data matching and prediction

In this subsection, it is assumed that only the hydraulic heads of the first 30 time-steps are available as observation data, and the hydraulic heads of the last 20 time-steps are utilized for testing the prediction from the estimated model parameters. Consider the case in subsection 6.2, and the inversion results of three surrogate-based methods are shown in **Figure 11** and **Table 3**. It can be seen that the estimated $\ln K$ field can capture the main patterns of the reference. However, the estimation is less accurate compared with the results in subsections 6.2-6.4 because less measurements are available. Indeed, high efficiency can be achieved with the proposed methods, which can be seen from **Table 3**.

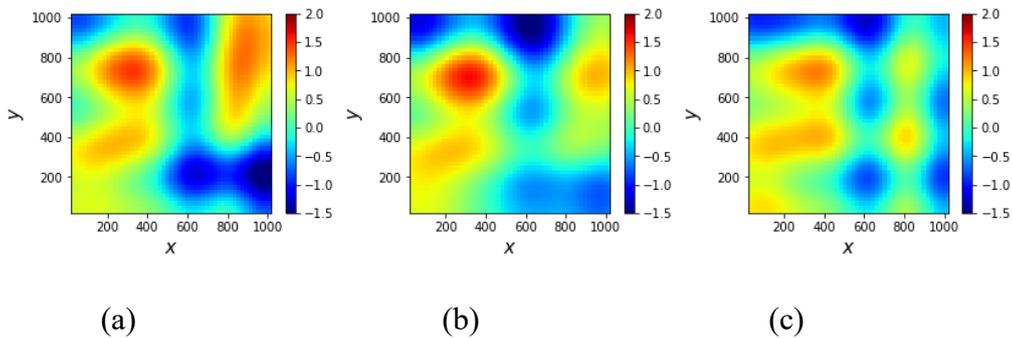

(a)                                    (b)                                 (c)

**Figure 11.** Inversion results of three surrogate-based methods: (a) IES; (b) gradient method; (c) and training method.

**Table 3.** Inversion results of three surrogate-based methods.

| | Realization Number | RMSE | Time (s) |
| --- | --- | --- | --- |
| IES | 100 | 0.636 | 4.7468 |
| Gradient | 1 | 0.776 | 0.1454 |
| Training | 1 | 0.719 | 6.8519 |

The data matching and prediction results of points 1 and 2 are shown in **Figure 12**, in which the blue dashed lines indicate the beginning of prediction. The data matching and prediction results of remaining points are shown in **Appendix A**. It can be seen that the hydraulic heads from the updated model parameters can match the reference well, even in the prediction period. The standard deviation at the initial and final steps for the IES method are presented in **Figure 13**. It can be seen that the final standard deviation of the ensemble increases along the *x* direction due to the fact that the flow moves from left to right, and the hydraulic heads at points 2 and 4 indicated in **Figure 12** and **Appendix A**, respectively, change slowly for the first 30 time-steps. Therefore, less information can be obtained from hydraulic head measurements from points 2 and 4, which leads to greater uncertainty.

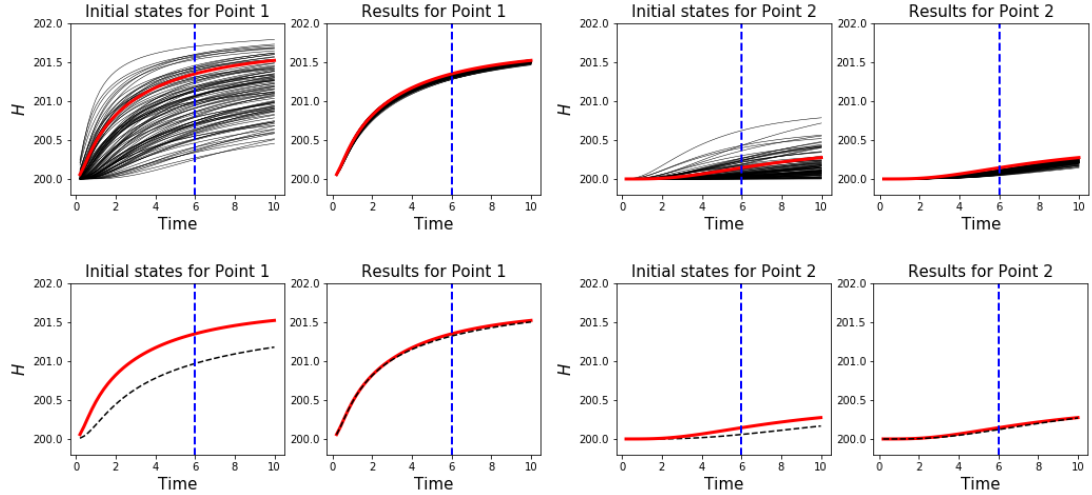

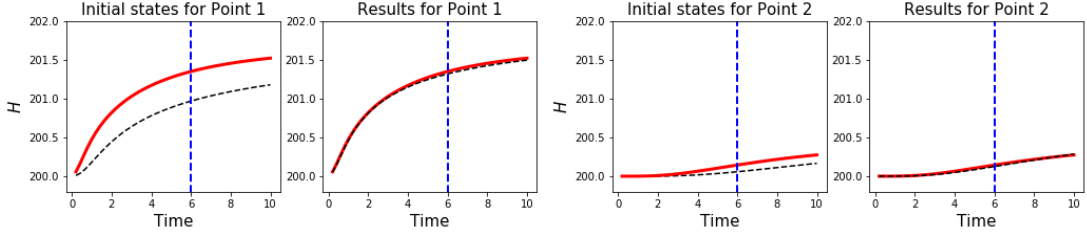

**Figure 12.** Data matching and prediction results of points 1 and 2 with different methods: IES method (first row); gradient method (second row); and training method (third row).

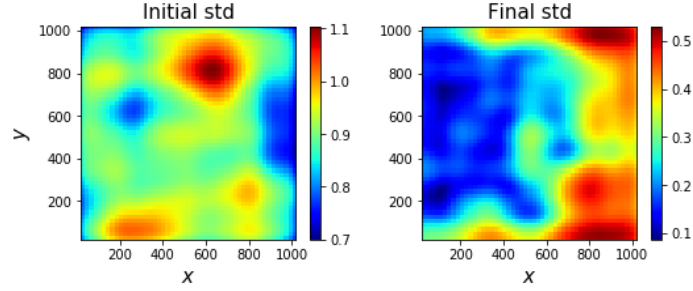

**Figure 13.** Standard deviation at initial and final steps with the IES method.

**6.6 Inversion for high-resolution field**

In this case, a high-resolution conductivity field is used as a reference. When generating the reference field with KLE, 90% of energy is retained to maintain more specific information of the conductivity field, which leads to 60 terms in the expansion. The high-resolution reference field is shown in **Figure 14(a)**. The TgNN surrogate previously constructed for the low-resolution conductivity field is employed, in which 20 independent random variables are used to parameterize the random field. Here, we aim to test the performance of the proposed methods for estimating the high-resolution field when the TgNN surrogate is constructed for the low-resolution field.

Three TgNN surrogate-based methods are implemented for this case, with $N_e = 100$, 1, and 1, respectively. The estimation results of high-resolution $\ln K$ with different methods are shown in **Figure 14** and **Table 4**. From the figure, it can be seen that, although the estimation for detailed features is not sufficiently accurate, the major pattern of the high-resolution reference $\ln K$ field has been captured. Therefore, the proposed methods can be used to roughly estimate the property fields of the complicated subsurface formation. In fact, compared with the reference field, the underlying retained percentage of the correlation structure information is intentionally reduced in this case to test the robustness of the proposed surrogate-based methods. The influence of other prior statistical parameters will be explored further in section 6.9.

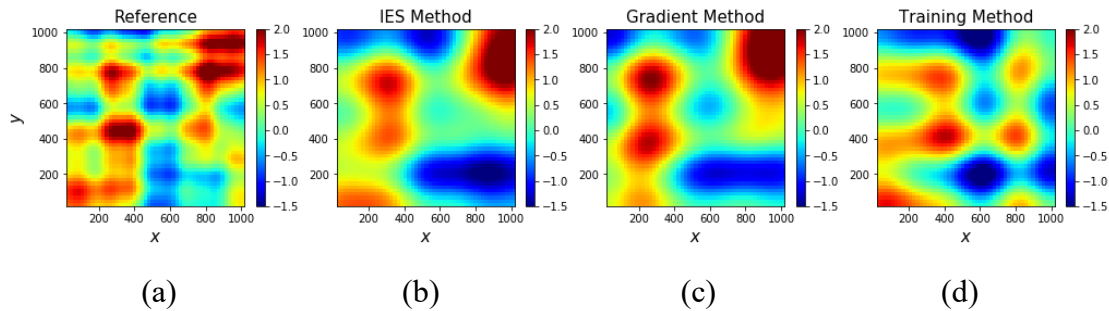

|     | (a) | (b) | (c) | (d) |

**Figure 14.** The high-resolution reference field (a) and the estimation results of high-resolution field with different methods: (b) IES; (c) gradient method; and (d) training method.

**Table 4.** Inversion results for high-resolution reference with three surrogate-based methods.

|  | Realization Number | RMSE | Time (s) |
| --- | --- | --- | --- |
| IES | 100 | 0.727 | 4.9113 |
| Gradient | 1 | 0.639 | 0.1713 |
| Training | 1 | 0.772 | 3.5333 |

## 6.7 Direct TgNN-geo inversion method

In this subsection, the performance of the proposed direct TgNN-geo inversion method is tested. The case introduced in subsection 6.2 is considered here and the difference is that, apart from the measurements of hydraulic heads, the direct measurements of hydraulic conductivity are assumed to be available at the observation points. Requiring a number of hydraulic conductivity measurements constitutes a disadvantage of the PINN-no-geo method for inverse modeling. Therefore, the proposed TgNN-geo method attempts to alleviate the problem by incorporating geostatistical information of the hydraulic conductivity field.

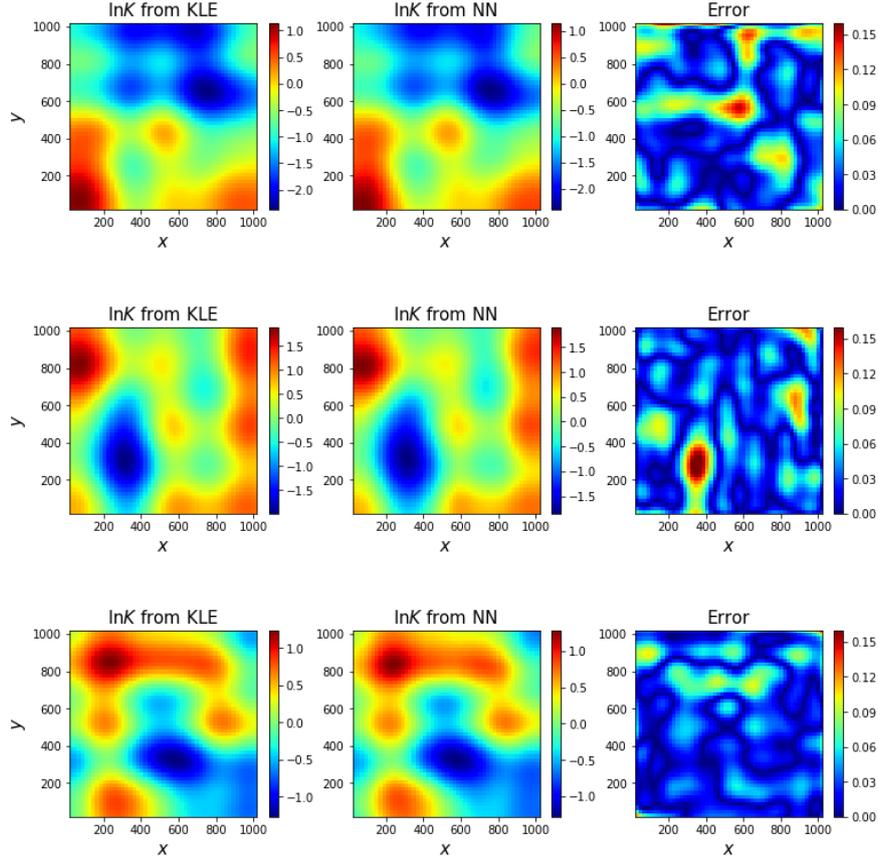

**Figure 15.** The testing results for three randomly generated realizations.

300 hydraulic conductivity fields are generated with KLE, which honor the first two moments of the random field, as introduced in subsection 6.2. The generated hydraulic conductivity fields constitute a training dataset, and are used to train a neural network, $\hat{K}(\mathbf{x}, \xi; \theta_{para}) = NN_{para}(\mathbf{x}, \xi; \theta_{para})$. The training process takes approximately 22.95 min (1377.17 s) for 2000 epochs with learning rate of 0.001. The testing results of trained $NN_{para}(\mathbf{x}, \xi; \theta_{para})$ for three randomly generated realizations are shown in **Figure 15**. It can be seen that the performance of the trained neural network is satisfactory, which has learned the geostatistical information from KLE. Then, $NN_{para}(\mathbf{x}, \xi; \theta_{para})$ can be used to couple with the network $\hat{h}(\mathbf{x}, t; \theta_h) = NN_h(\mathbf{x}, t; \theta_h)$ for inverse modeling via the deep-learning framework. The parameters $\theta_{para}$ of neural network $NN_{para}(\mathbf{x}, \xi; \theta_{para})$ are fixed, and the inputs $\xi$ are tuned during the inversion process. Consequently, the number of parameters to be trained is reduced significantly compared with the PINN-no-geo method.

The measurements of hydraulic head and conductivity at the five observation points indicated in **Figure 2(b)** are used to constitute the data mismatch loss, as illustrated in the last two terms of Eq. (33). The physics-based loss is also incorporated, as shown in the first three terms of Eq. (33). 10,000 collocation points are employed to enforce the physics constraints. 2,000 epochs of training are performed with learning rate of 0.001. The final estimated result with TgNN-geo, as well as the corresponding error, are presented in **Figure 16** and **Table 5**, and the estimated result with PINN-no-geo is also shown for comparison. It is obvious that the estimation from TgNN-geo is superior to that from PINN-no-geo given the same number of measurements. It can also be seen that the PINN-no-geo has a higher degree of freedom without constraining with any geostatistical information to capture the spatial correlation during the training process, and produces unsatisfactory estimation with sparse measurements. Moreover, the consumed time for direct-deep-learning-inversion is significantly less than that for the deep-learning surrogate construction, as shown in **Table 5**. This is because the TgNN surrogate is constructed for uncertain model parameters, which can predict SPDE solutions for different random fields, while the direct TgNN-geo method is used for a specific target random field and a particular problem setup.

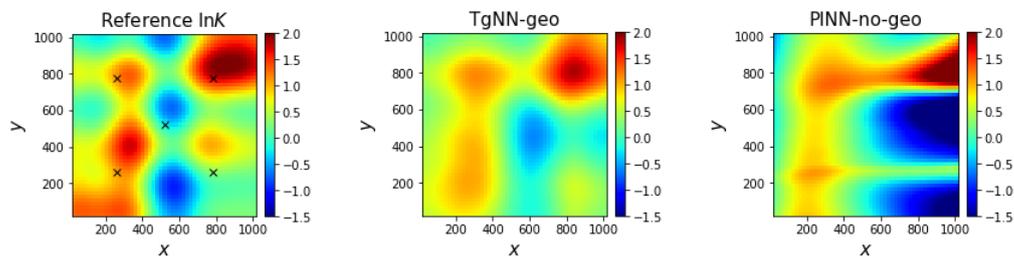

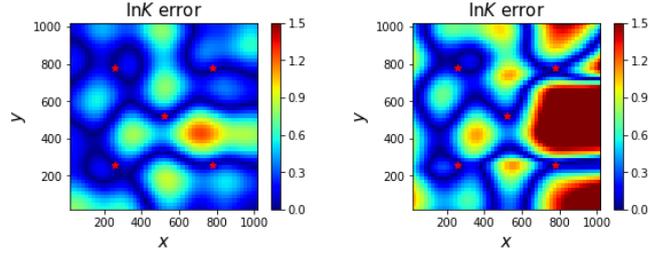

**Figure 16.** Estimation results and errors with TgNN-geo and PINN-no-geo method.

**Table 5.** Inversion results with TgNN-geo and PINN-no-geo methods when hydraulic conductivity measurements are available.

|  | Observation Points | Collocation Points | Epochs | Training Time (s) | RMSE |
| --- | --- | --- | --- | --- | --- |
| TgNN-geo | 5 | 10000 | 2000 | 290.566 | 0.423 |
| PINN-no-geo | 5 | 10000 | 2000 | 305.597 | 0.897 |

### 6.8 Methods comparison

The performances of the proposed methods are compared in this subsection. In order to compare the two categories of methods for inverse modeling under the same conditions, two types of scenarios are considered. The first one considers the case in subsection 6.7, in which both indirect (i.e., hydraulic head) and direct (i.e., hydraulic conductivity) measurements are available at the observation points. In the second case, the direct deep-learning inversion methods, as well as deep-learning surrogate-based methods, are implemented with only the hydraulic head measurements. Furthermore, in order to investigate the effect of the number of observation points, different observation location settings are studied for all of the five algorithms discussed in this work.

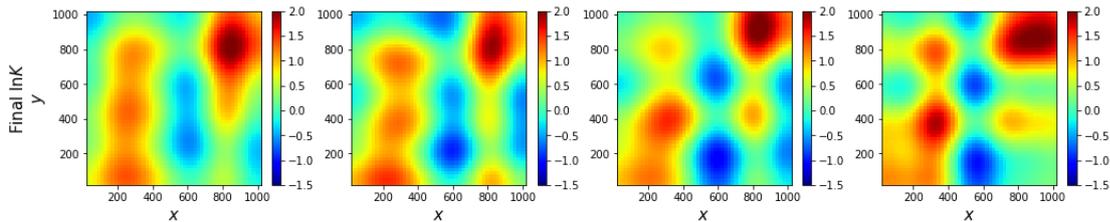

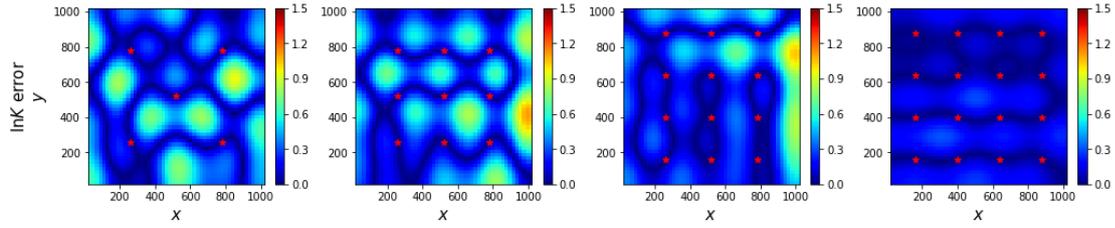
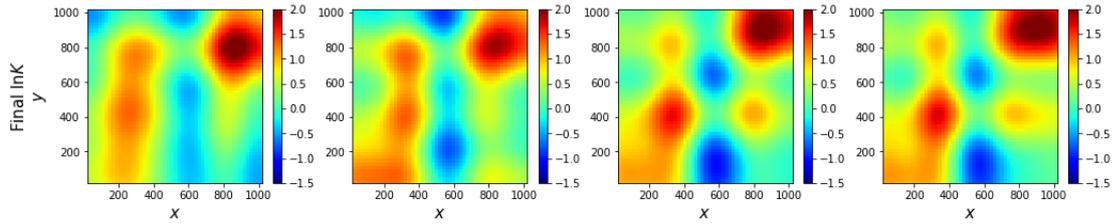

(a) TgNN surrogate-based IES

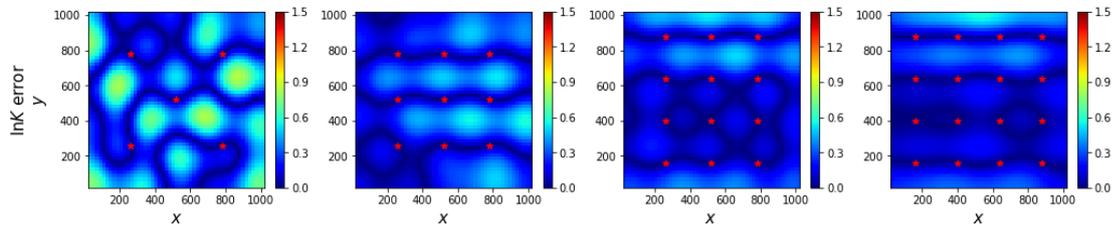
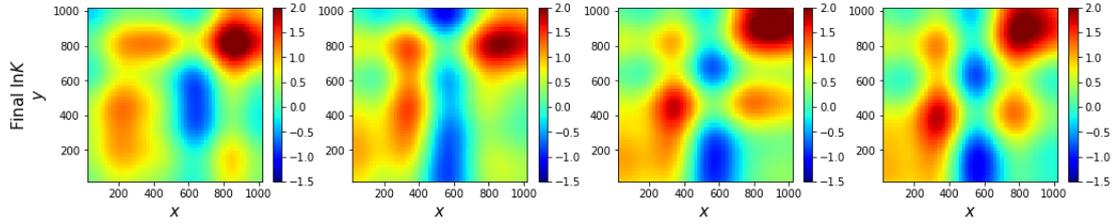

(b) TgNN surrogate-based gradient method

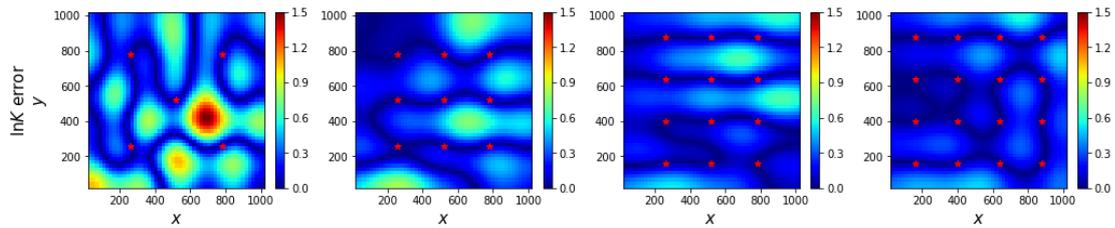
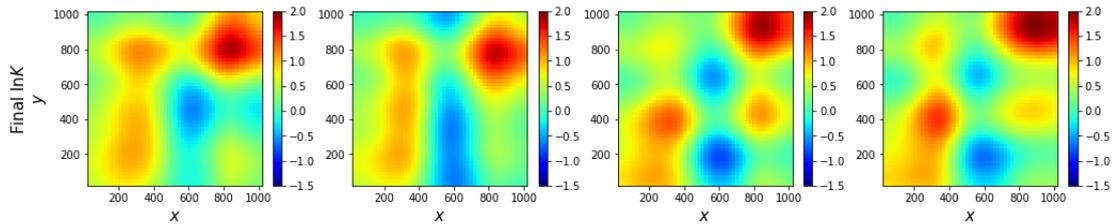

(c) TgNN surrogate-based training method

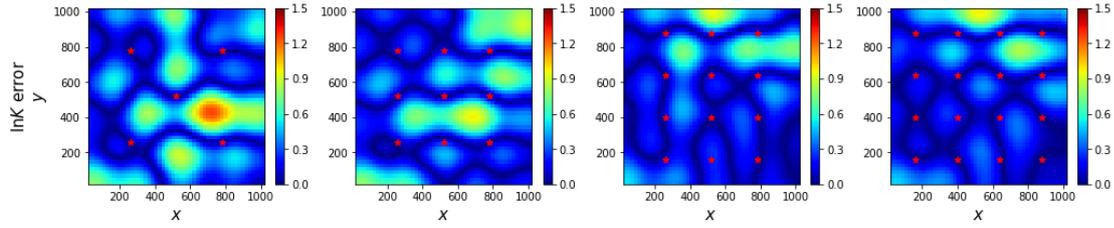

(d) Direct TgNN-geo inversion method

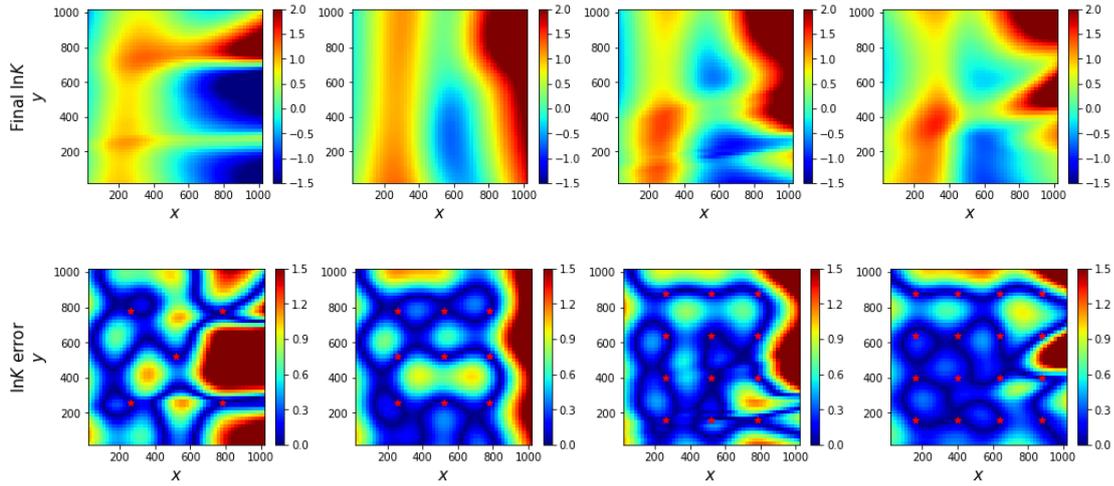

(e) Direct PINN-no-geo inversion method

**Figure 17.** Estimation results of the five methods with different numbers of both hydraulic head and conductivity measurements. The red dots in the second panel in each case indicate the measurement locations.

The results of the five methods with different numbers of available hydraulic head and conductivity measurements are shown in **Figure 17, Figure 18(a)**, and **Table 6**. The results without hydraulic conductivity measurements are presented in **Figure 18(b).** From these figures, it can be seen that the results with both hydraulic head and conductivity measurements are more accurate than those without hydraulic conductivity measurements. In addition, the performance gets better as the number of observation points increases. Furthermore, when the hydraulic conductivity measurements are not available, both the TgNN-geo and PINN-no-geo perform poorly (**Figure 18(b)**), especially the PINN-no-geo method.

Compared with TgNN-geo, the TgNN surrogate-based IES method, the gradient method, and the training method can provide equivalent accuracy given the same number of hydraulic head and conductivity measurements, as shown in **Figure 17** and **Figure 18(a)**. Considering the time consumed to construct the TgNN surrogate (7,857.762 s), however, the direct TgNN-geo inversion method is more efficient, which doesn't require a pre-constructed surrogate. However, once the TgNN surrogate is constructed, the inversion processes takes a very small amount of time, as shown in **Table 6**, and the trained surrogate can be directly used for cases with different setups (e.g., variance, correlation scale), or different types and amount of measurement data. Furthermore, the direct TgNN-geo inversion method only works well when direct hydraulic conductivity measurements are available, while the TgNN surrogate-based methods do not require this. It can also be seen that both the proposed TgNN surrogate-based methods and the direct TgNN-geo inversion method perform much better than the PINN-no-geo method under sparse measurement data.

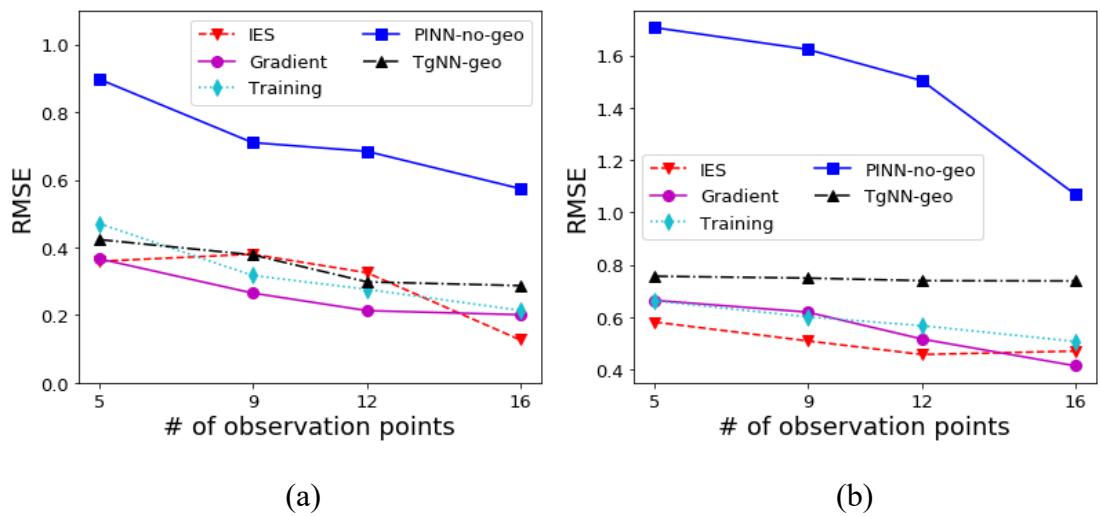

(a) (b)

**Figure 18.** RMSE of five methods for different numbers of observations: (a) when both hydraulic head and conductivity measurements are available; (b) when only hydraulic head measurements are available.

**Table 6.** RMSE and inversion consumed time of the five methods with different numbers of both hydraulic head and conductivity measurements.

|  | Time (s) | RMSE | Time (s) | RMSE | Time (s) | RMSE | Time (s) | RMSE |
|---|---|---|---|---|---|---|---|---|
| Observation Points | 5 | | 9 | | 12 | | 16 | |
| IES | 2.4230 | 0.359 | 4.7845 | 0.380 | 6.9057 | 0.325 | 11.3170 | 0.127 |
| Gradient | 0.0796 | 0.366 | 0.0817 | 0.265 | 0.0866 | 0.213 | 0.0966 | 0.201 |
| Training | 11.3061 | 0.470 | 11.7067 | 0.317 | 11.7149 | 0.276 | 11.5279 | 0.214 |
| TgNN-geo | 290.5658 | 0.423 | 326.8532 | 0.378 | 314.5828 | 0.298 | 347.8243 | 0.287 |
| PINN-no-geo | 305.5972 | 0.897 | 308.5481 | 0.710 | 322.5508 | 0.684 | 328.9147 | 0.574 |

## 6.9 Influence of prior statistics

Considering that the prior statistics of the model parameter field may not be accurately known in engineering practice, in this subsection, the influence of the prior statistics of the model parameter field, such as variance and correlation length, are investigated. Three hydraulic conductivity fields with $\sigma^2_{\ln K} = 0.5$, $\sigma^2_{\ln K} = 2.0$, and $\eta_x = \eta_y = 204[L]$, respectively, are chosen as reference fields for three new cases, as shown in **Figure 19(a)**. The color bars are set to be the same to highlight the different spatial variability of the reference fields. Other statistical information remains the same as the case in subsection 6.2. 16 observation points are selected, where the hydraulic head and/or conductivity measurements are collected, as shown in **Figure 19(b)**.

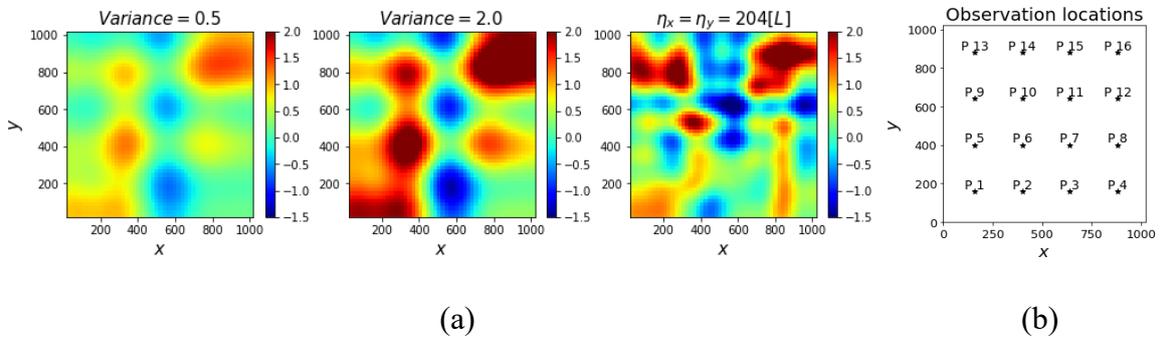

(a) (b)

**Figure 19.** Reference fields with different variance and correlation length (a) and the location of observation points (b).

### 6.9.1 Influence of prior statistics for TgNN surrogate-based methods

The three new cases are investigated with TgNN surrogate methods in this subsection. The TgNN surrogate is still constructed with $\sigma^2_{\ln K} = 1.0$ and $\eta_x = \eta_y = 408[L]$ as previously. Firstly, the three cases are implemented with only the hydraulic head measurements, and the results of the cases with the three TgNN surrogate-based methods are shown in **Figure 20** and **Table 7**. It can be seen that the imprecise variance of the field can be easily corrected during the inversion process, and the degree of spatial variability is captured, as shown in **Figure 20**. The imprecise correlation length has a more obvious impact on the performance, since the TgNN surrogate is constructed with preset limited KLE terms (20 terms), and it is difficult to totally recover the high-resolution field, as discussed in subsection 6.6. However, the results are still satisfactory, and present similar general patterns to the reference field. From these cases, it can be seen that the TgNN surrogate-based methods are effective with imprecise variance and correlation length of random model parameter field in a certain range.

Moreover, the three cases are implemented with both the hydraulic head and conductivity measurements. The results are shown in **Figure 21** and **Table 7.** It is obvious that the results are much better than the cases with only hydraulic head measurements. In addition, when both the hydraulic head and conductivity measurements are available, a similar conclusion can be drawn that the imprecise variance of the field can be more readily corrected than the correlation length.

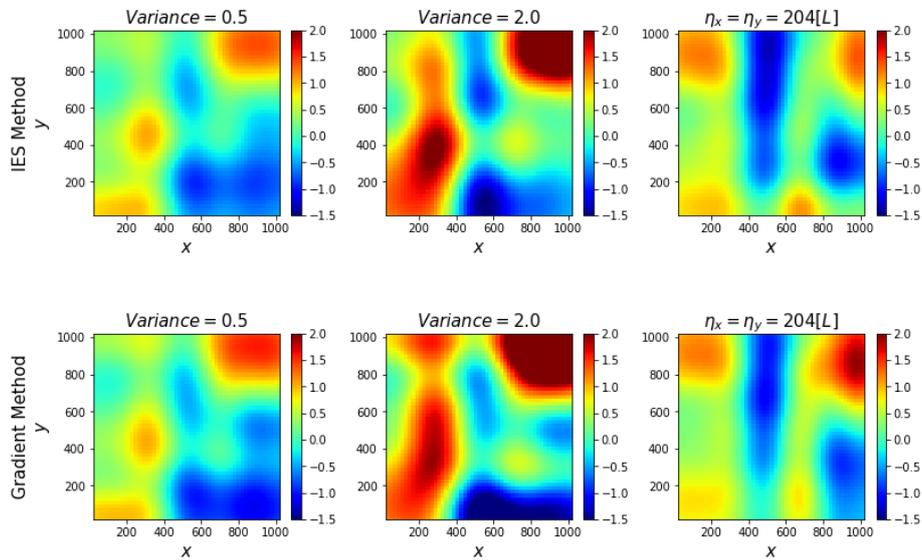

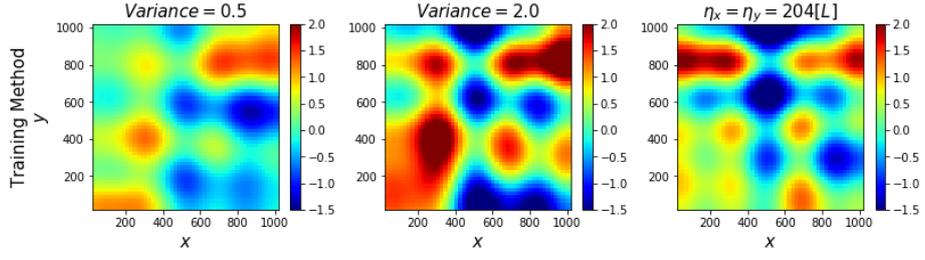

**Figure 20.** Results for the three new cases ($\sigma_{\ln K}^2 = 0.5$, $\sigma_{\ln K}^2 = 2.0$, $\eta_x = \eta_y = 204[L]$) with TgNN surrogate-based methods when only hydraulic head measurements are available.

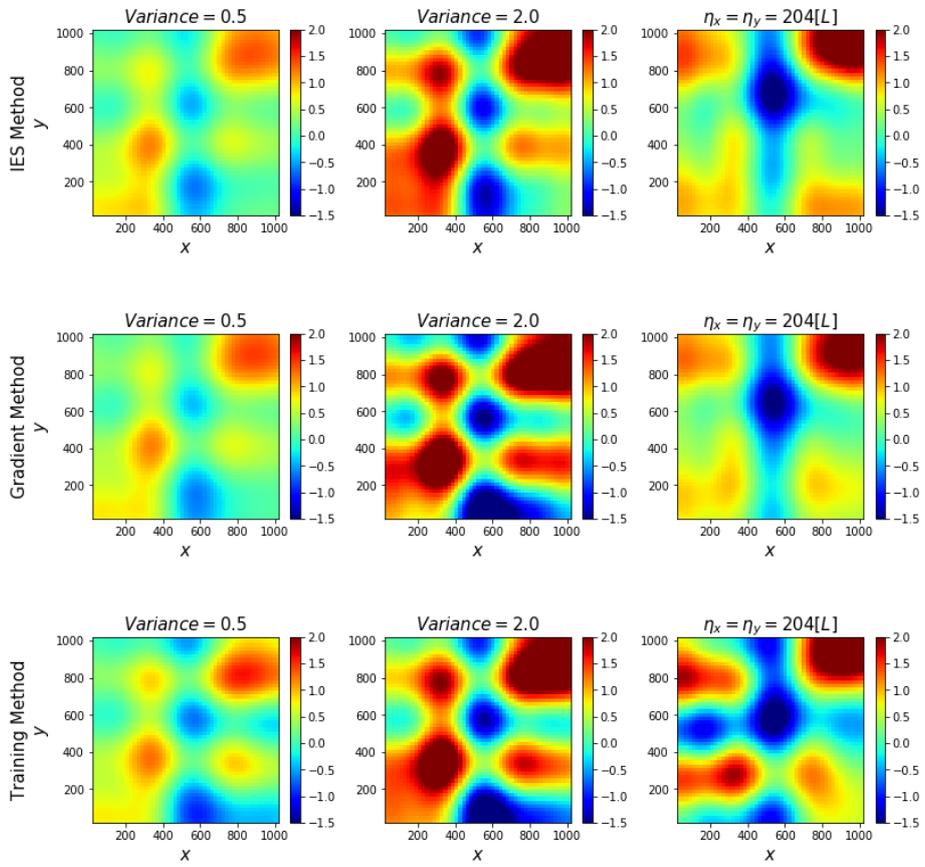

**Figure 21.** Results for the three new cases ($\sigma_{\ln K}^2 = 0.5$, $\sigma_{\ln K}^2 = 2.0$, $\eta_x = \eta_y = 204[L]$) with TgNN surrogate-based methods when both hydraulic head and conductivity measurements are available.

**Table 7.** RMSE of the five methods with imprecise prior statistics.

|  | $\sigma^2_{\ln K} = 0.5$ | $\sigma^2_{\ln K} = 2.0$ | $\eta_x = \eta_y = 204[L]$ |
|---|---|---|---|
| IES (with only *h* measurements) | 0.434 | 0.538 | 0.815 |
| Gradient (with only *h* measurements) | 0.488 | 0.802 | 0.763 |
| Training (with only *h* measurements) | 0.436 | 0.705 | 0.838 |
| IES (with both *h* and ln*K* measurements) | 0.090 | 0.231 | 0.666 |
| Gradient (with both *h* and ln*K* measurements) | 0.138 | 0.576 | 0.624 |
| Training (with both *h* and ln*K* measurements) | 0.209 | 0.486 | 0.747 |
| TgNN-geo (with both *h* and ln*K* measurements) | 0.220 | 0.432 | 0.620 |
| PINN-no-geo (with both *h* and ln*K* measurements) | 0.401 | 0.644 | 0.860 |

### 6.9.2 Influence of prior statistics for direct-deep-learning-inversion methods

The influence of prior statistics for TgNN-geo and PINN-no-geo is tested here with the three new cases. In this subsection, not only the hydraulic head, but also the hydraulic conductivity measurements at the observation points, are assumed to be available to improve the performance of the two direct-deep-leaning-inversion methods, as indicated by the results in subsection 6.8. The results of the three new cases with the two direct-deep-leaning-inversion methods are shown in **Figure 22** and **Table 7**. Here, it is important to note that the PINN-no-geo is not affected by imprecise geostatistical information because this information is not considered in the PINN-no-geo method; whereas, the TgNN-geo is affected by imprecise geostatistical information, which is incorporated when approximating the hydraulic conductivity. Similar to the results in subsection 6.9.1, the imprecise variance of the random field is easily corrected, while the imprecise correlation length is more difficult to amend since the random dimension is already settled when constructing the neural network $\hat{K}(\mathbf{x},\xi;\ \theta_{para}) = NN_{para}(\mathbf{x},\xi;\ \theta_{para})$. However, compared with PINN-no-geo, TgNN-geo still performs better in these cases, as shown in **Figure 22** and **Table 7**, which means that the available imprecise geostatistical information is still beneficial. Moreover, when the geostatistical information is largely biased, it may be difficult to achieve satisfactory results using TgNN-geo. Compared with the results of TgNN surrogate-based methods when both the hydraulic head and conductivity measurements are available, one can see that TgNN-geo achieves equivalent performance without any surrogate. In addition, both the TgNN surrogate methods and TgNN-geo outperform the PINN-no-geo method with imprecise prior statistics.

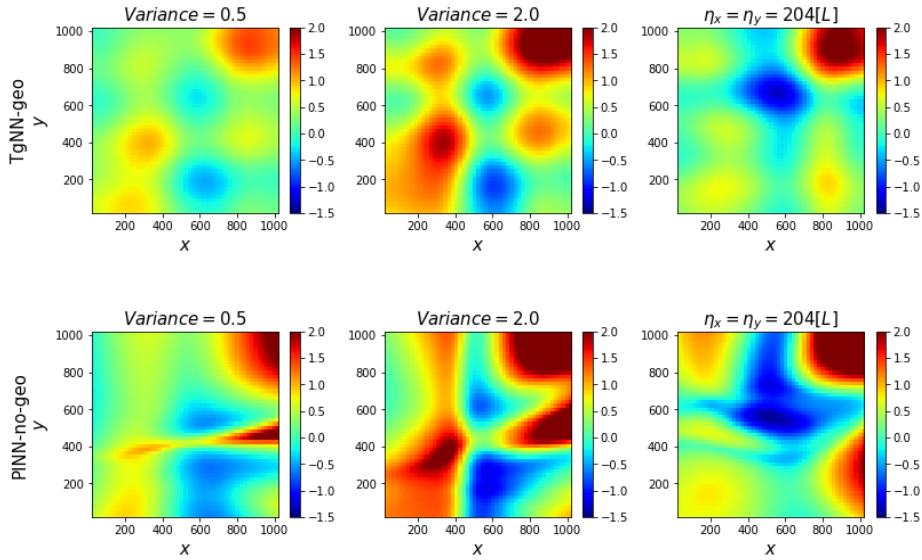

**Figure 22.** Results for the three new cases ($\sigma_{\ln K}^2 = 0.5$, $\sigma_{\ln K}^2 = 2.0$, $\eta_x = \eta_y = 204[L]$) with TgNN-geo and PINN-no-geo methods.

## 7 Discussions and conclusions

In this work, two categories of inversion methods are introduced and compared. The first category is deep-learning surrogate-based methods, including TgNN surrogate-based IES, TgNN surrogate-based gradient method, and TgNN surrogate-based training method. The latter two methods take advantage of the differentiable property of neural networks. The second category is the direct-deep-learning-inversion method, in which the TgNN-geo is proposed, which can incorporate geostatistical information. Several two-dimensional subsurface flow problems are designed to test the performance of the proposed methods.

For the TgNN surrogate-based methods, the TgNN surrogate is trained by matching the available simulation data and honoring physical/engineering principles at selected collocation points. The TgNN surrogate can be constructed with limited labeled data, or even in a label-free manner, which is an advantage of the TgNN surrogate as studied in Wang et al. (2020a). Although the training process of the TgNN surrogate may require some computational cost, the trained surrogate accelerates the inversion procedure significantly and can be used for solving new cases, such as varying prior statistics, and different spatial and temporal observations. Furthermore, the differentiability of the TgNN surrogate makes it possible to efficiently use the gradient method and training method for inversion tasks.

The performance of the three TgNN surrogate-based methods are first tested with a two-dimensional dynamic subsurface flow case, and then more complicated tasks, such as extrapolation to future times, high-resolution field estimation, and inversion under imprecise geostatistical information situations, are investigated. Satisfactory results demonstrate the robustness of the proposed methods.

***The gradient method*** needs a sensitivity matrix, which can be obtained from the TgNN surrogate. Updating only one realization is feasible using this method, which is much more efficient compared with ensemble-based methods. However, the performance of the gradient method is easily influenced by the initial guess of model parameter, and the uncertainty of the posterior cannot by quantified with only one realization. Nonetheless, the gradient method can be implemented in an ensemble manner, which can assist to improve stability and quantify estimation uncertainty. ***The IES method*** does not need explicit computation of the sensitivity matrix, and instead the covariance calculated from the ensemble is utilized. In this method, a group of realizations need to be updated simultaneously, and thus the uncertainty of posterior can be quantified, and the initialization of model parameter has little effect on performance. Moreover, the iterative update of realizations in the ensemble can be implemented efficiently with the TgNN surrogate. ***The training method*** optimizes the objective function of inverse problems directly with embedded algorithms of deep-learning frameworks, such as Adam embedded in Pytorch. Therefore, the training method can be operated easily. Similar to the gradient method, only one realization is needed to update, and its feasibility benefits from the differentiable property of the TgNN surrogate.

The PINN-no-geo method has been utilized for inverse modeling in the past. However, the requirement of a large number of direct model parameter measurements (e.g., hydraulic conductivity) and lack of geostatistical information constraint constitute the major drawbacks of this method. The proposed ***TgNN-geo method*** deals with this problem by incorporating the geostatistical information of the random field. In TgNN-geo, two neural networks are introduced to approximate the random model parameter and the solution, respectively. In order to honor prior geostatistical information of the random model parameter, the neural network for approximating the random model parameter is trained by using the realizations generated from KLE. By minimizing the loss function of TgNN-geo, estimation of the model parameter and approximation of the model solution can be simultaneously obtained. By learning the geostatistical information from KLE, the TgNN-geo works well, even under conditions of sparse observation points. Although some geostatistical information may be imprecise, the available partial information is still beneficial. The hydraulic conductivity measurements play an important role for TgNN-geo, and without this kind of measurement, performance is unsatisfactory.

The TgNN surrogate-based methods can also be implemented with hydraulic conductivity measurements for comparison with the direct TgNN-geo inversion method, and equivalent performance can be achieved. The advantage of the direct TgNN-geo inversion method is that no surrogates are needed when solving inverse problems, which can save some computational cost. Regarding the requirement of the direct measurements of the model parameters being inferred (e.g., hydraulic conductivity), the TgNN surrogate-based methods can work well without this kind of measurement, which constitutes an advantage of this category of methods.

**Acknowledgements**

This work is partially funded by the National Natural Science Foundation of China (Grant No. 51520105005) and the National Science and Technology Major Project of China (Grant No. 2017ZX05009-005 and 2017ZX05049-003). All of the data of results shown in this paper are made available for download from a public repository of research data through the following

link:

https://figshare.com/articles/dataset/Data_for_Deep_Learning_based_Inverse_Modeling_Approaches/12587048 (DOI: 10.6084/m9.figshare.12587048).

**References**

Abadi, M., Barham, P., Chen, J., Chen, Z., Davis, A., Dean, J., et al. (2016). *Tensorflow: A system for large-scale machine learning*. Paper presented at the 12th Symposium on Operating Systems Design and Implementation.

Adler, J., & Öktem, O. (2017). Solving ill-posed inverse problems using iterative deep neural networks. *Inverse Problems, 33*(12), 124007. https://doi.org/10.1088/1361-6420/aa9581

Anterion, F., Eymard, R., & Karcher, B. (1989). *Use of Parameter Gradients for Reservoir History Matching*. Paper presented at the SPE Symposium on Reservoir Simulation, Houston, Texas. https://doi.org/10.2118/18433-MS

Antholzer, S., Haltmeier, M., & Schwab, J. (2019). Deep learning for photoacoustic tomography from sparse data. *Inverse Problems in Science and Engineering, 27*(7), 987-1005. https://doi.org/10.1080/17415977.2018.1518444

Arridge, S. R. (1999). Optical tomography in medical imaging. *Inverse Problems, 15*(2), R41-R93. http://dx.doi.org/10.1088/0266-5611/15/2/022

Bottou, L. (2010). Large-scale machine learning with stochastic gradient descent. In *Proceedings of COMPSTAT'2010* (pp. 177-186): Springer.

Bunks, C., Saleck, F. M., Zaleski, S., & Chavent, G. (1995). Multiscale seismic waveform inversion. *Geophysics, 60*(5), 1457-1473. https://doi.org/10.1190/1.1443880

Canchumuni, S. W. A., Emerick, A. A., & Pacheco, M. A. C. (2019). Towards a robust parameterization for conditioning facies models using deep variational autoencoders and ensemble smoother. *Computers & Geosciences, 128*, 87-102. https://doi.org/10.1016/j.cageo.2019.04.006

Carrera, J., & Neuman, S. P. (1986a). Estimation of Aquifer Parameters Under Transient and Steady State Conditions: 1. Maximum Likelihood Method Incorporating Prior Information. *Water Resources Research, 22*(2), 199-210. https://doi.org/10.1029/WR022i002p00199

Carrera, J., & Neuman, S. P. (1986b). Estimation of Aquifer Parameters Under Transient and Steady State Conditions: 2. Uniqueness, Stability, and Solution Algorithms. *Water Resources Research, 22*(2), 211-227. https://doi.org/10.1029/WR022i002p00211

Carrera, J., & Neuman, S. P. (1986c). Estimation of Aquifer Parameters Under Transient and Steady State Conditions: 3. Application to Synthetic and Field Data. *Water Resources Research, 22*(2), 228-242. https://doi.org/10.1029/WR022i002p00228

Chan, S., & Elsheikh, A. H. (2019). Parametric generation of conditional geological realizations using generative neural networks. *Computational Geosciences, 23*(5), 925-952. https://doi.org/10.1007/s10596-019-09850-7

Chang, H., Chen, Y., & Zhang, D. (2010). Data assimilation of coupled fluid flow and geomechanics using the ensemble kalman filter. *SPE Journal, 15*(2), 382-394. https://doi.org/10.2118/118963-PA

Chang, H., Liao, Q., & Zhang, D. (2017). Surrogate model based iterative ensemble smoother for subsurface flow data assimilation. *Advances in Water Resources, 100*, 96-108. http://dx.doi.org/10.1016/j.advwatres.2016.12.001

Chavent, G., Dupuy, M., & Lemmonier, P. (1975). History Matching by Use of Optimal Theory. *SPE Journal, 15*(01), 74-86. https://doi.org/10.2118/4627-PA

Chen, W. H., Gavalas, G. R., Seinfeld, J. H., & Wasserman, M. L. (1974). A New Algorithm for Automatic History Matching. *SPE Journal, 14*(06), 593-608. https://doi.org/10.2118/4545-PA

Chen, Y., & Oliver, D. S. (2012). Ensemble Randomized Maximum Likelihood Method as an Iterative Ensemble Smoother. *Mathematical Geosciences, 44*(1), 1-26. https://doi.org/10.1007/s11004-011-9376-z

Chen, Y., & Oliver, D. S. (2013). Levenberg–Marquardt forms of the iterative ensemble smoother for efficient history matching and uncertainty quantification. *Computational Geosciences, 17*(4), 689-703. https://doi.org/10.1007/s10596-013-9351-5

Chen, Y., & Zhang, D. (2006). Data assimilation for transient flow in geologic formations via ensemble Kalman filter. *Advances in Water Resources, 29*(8), 1107-1122. https://doi.org/10.1016/j.advwatres.2005.09.007

Evensen, G. (1994). Sequential data assimilation with a nonlinear quasi‐geostrophic model using Monte Carlo methods to forecast error statistics. *Journal of Geophysical Research: Oceans, 99*(C5), 10143-10162. https://doi.org/10.1029/94JC00572

Ghanem, R. G., & Spanos, P. D. (2003). *Stochastic finite elements: a spectral approach*. New York: Dover Publications.

Gu, Y., & Oliver, D. S. (2005). History Matching of the PUNQ-S3 Reservoir Model Using the Ensemble Kalman Filter. *SPE Journal, 10*(02), 217-224. https://doi.org/10.2118/89942-PA

Gu, Y., & Oliver, D. S. (2007). An Iterative Ensemble Kalman Filter for Multiphase Fluid Flow Data Assimilation. *SPE Journal, 12*(04), 438-446. https://doi.org/10.2118/108438-PA

He, Q., Barajas-Solano, D., Tartakovsky, G., & Tartakovsky, A. M. (2020). Physics-informed neural networks for multiphysics data assimilation with application to subsurface transport. *Advances in Water Resources, 141*, 103610. https://doi.org/10.1016/j.advwatres.2020.103610

Houtekamer, P. L., & Mitchell, H. L. (2001). A sequential ensemble Kalman filter for atmospheric data assimilation. *Monthly Weather Review, 129*(1), 123-137. https://doi.org/10.1175/1520-0493(2001)129<0123:ASEKFF>2.0.CO;2

Houtekamer, P. L., Mitchell, H. L., Pellerin, G., Buehner, M., Charron, M., Spacek, L., & Hansen, B. (2005). Atmospheric data assimilation with an ensemble Kalman filter: Results with real observations. *Monthly Weather Review, 133*(3), 604-620. https://doi.org/10.1175/MWR-2864.1

Jin, K. H., McCann, M. T., Froustey, E., & Unser, M. (2017). Deep Convolutional Neural

Network for Inverse Problems in Imaging. *IEEE Transactions on Image Processing, 26*(9), 4509-4522. https://doi.org/10.1109/TIP.2017.2713099

Ju, L., Zhang, J., Meng, L., Wu, L., & Zeng, L. (2018). An adaptive Gaussian process-based iterative ensemble smoother for data assimilation. *Advances in Water Resources, 115*, 125-135. https://doi.org/10.1016/j.advwatres.2018.03.010

Kingma, D. P., & Ba, J. L. (2015). *Adam: A Method for Stochastic Optimization*. Paper presented at the International conference on learning representations.

Laloy, E., Hérault, R., Jacques, D., & Linde, N. (2018). Training‐Image Based Geostatistical Inversion Using a Spatial Generative Adversarial Neural Network. *Water Resources Research, 54*(1), 381-406. https://doi.org/10.1002/2017WR022148

Laloy, E., Hérault, R., Lee, J., Jacques, D., & Linde, N. (2017). Inversion using a new low-dimensional representation of complex binary geological media based on a deep neural network. *Advances in Water Resources, 110*, 387-405. https://doi.org/10.1016/j.advwatres.2017.09.029

Le, D. H., Emerick, A. A., & Reynolds, A. C. (2016). An Adaptive Ensemble Smoother With Multiple Data Assimilation for Assisted History Matching. *SPE Journal, 21*(06), 2195-2207. https://doi.org/10.2118/173214-PA

Li, R., Reynolds, A. C., & Oliver, D. S. (2003). Sensitivity Coefficients for Three-Phase Flow History Matching. *Journal of Canadian Petroleum Technology, 42*(04), 70-77. https://doi.org/10.2118/03-04-04

Li, S., Liu, B., Ren, Y., Chen, Y., Yang, S., Wang, Y., & Jiang, P. (2020). Deep-Learning Inversion of Seismic Data. *IEEE Transactions on Geoscience and Remote Sensing, 58*(3), 2135-2149. https://doi.org/10.1109/TGRS.2019.2953473

Mo, S., Zabaras, N., Shi, X., & Wu, J. (2019). Deep Autoregressive Neural Networks for High‐Dimensional Inverse Problems in Groundwater Contaminant Source Identification. *Water Resources Research, 55*(5), 3856-3881. https://doi.org/10.1029/2018WR024638

Nævdal, G., Johnsen, L. M., Aanonsen, S. I., & Vefring, E. H. (2005). Reservoir monitoring and continuous model updating using ensemble Kalman filter. *SPE Journal, 10*(1), 66-74. https://doi.org/10.2118/84372-MS

Oliver, D. S., Reynolds, A. C., & Liu, N. (2008). *Inverse Theory for Petroleum Reservoir Characterization and History Matching*. New York: Cambridge University Press.

Paszke, A., Gross, S., Chintala, S., Chanan, G., Yang, E., DeVito, Z., et al. (2017). *Automatic differentiation in pytorch*. Paper presented at the 31st Conference on Neural Information Processing Systems (NIPS 2017), Long Beach, CA, USA.

Raissi, M., Perdikaris, P., & Karniadakis, G. E. (2019). Physics-informed neural networks: A deep learning framework for solving forward and inverse problems involving nonlinear partial differential equations. *Journal of Computational Physics, 378*, 686-707. https://doi.org/10.1016/j.jcp.2018.10.045

Reichle, R. H., McLaughlin, D. B., & Entekhabi, D. (2002). Hydrologic data assimilation with the ensemble Kalman filter. *Monthly Weather Review, 130*(1), 103-114. https://doi.org/10.1175/1520-0493(2002)130<0103:HDAWTE>2.0.CO;2

Skjervheim, J.-a., & Evensen, G. (2011). *An Ensemble Smoother for Assisted History Matching*.

Paper presented at the SPE Reservoir Simulation Symposium, The Woodlands, Texas, USA. https://doi.org/10.2118/141929-MS

Tartakovsky, A. M., Marrero, C. O., Perdikaris, P., Tartakovsky, G. D., & Barajas‐Solano, D. (2020). Physics‐Informed Deep Neural Networks for Learning Parameters and Constitutive Relationships in Subsurface Flow Problems. *Water Resources Research, 56*(5). https://doi.org/10.1029/2019WR026731

Tavakoli, R., Tavakoli, R., Reynolds, A. C., & Reynolds, A. C. (2011). Monte Carlo simulation of permeability fields and reservoir performance predictions with SVD parameterization in RML compared with EnKF. *Computational Geosciences, 15*(1), 99-116. https://doi.org/10.1007/s10596-010-9200-8

Van Leeuwen, P. J., & Evensen, G. (1996). Data assimilation and inverse methods in terms of a probabilistic formulation. *Monthly Weather Review, 124*(12), 2898-2913. https://doi.org/10.1175/1520-0493(1996)124<2898:DAAIMI>2.0.CO;2

Wang, N., Chang, H., & Zhang, D. (2020a). Efficient Uncertainty Quantification for Dynamic Subsurface Flow with Surrogate by Theory-guided Neural Network. *arXiv preprint arXiv:.13560*.

Wang, N., Zhang, D., Chang, H., & Li, H. (2020b). Deep learning of subsurface flow via theory-guided neural network. *Journal of Hydrology, 584*, 124700. https://doi.org/10.1016/j.jhydrol.2020.124700

Wasserman, M. L., Emanuel, A. S., & Seinfeld, J. H. (1975). Practical Applications of Optimal-Control Theory to History-Matching Multiphase Simulator Models. *SPE Journal, 15*(04), 347-355. https://doi.org/10.2118/5020-PA

Wu, Y., & Lin, Y. (2020). InversionNet: An Efficient and Accurate Data-Driven Full Waveform Inversion. *IEEE Transactions on Computational Imaging, 6*, 419-433. https://doi.org/10.1109/TCI.2019.2956866

Wu, Z., Reynolds, A. C., & Oliver, D. S. (1999). Conditioning geostatistical models to two-phase production data. *SPE Journal, 4*(2), 142-155. https://doi.org/10.2118/56855-PA

Yang, P. H., & Watson, A. T. (1988). Automatic History Matching With Variable-Metric Methods. *SPE Reservoir Engineering, 3*(03), 995-1001. https://doi.org/10.2118/16977-PA

Zhang, D. (2001). *Stochastic methods for flow in porous media: coping with uncertainties*: Elsevier.

Zhang, D., & Lu, Z. (2004). An efficient, high-order perturbation approach for flow in random porous media via Karhunen–Loeve and polynomial expansions. *Journal of Computational Physics, 194*(2), 773-794. https://doi.org/10.1016/j.jcp.2003.09.015

Zhou, H., Gómez-Hernández, J. J., Hendricks Franssen, H.-J., & Li, L. (2011). An approach to handling non-Gaussianity of parameters and state variables in ensemble Kalman filtering. *Advances in Water Resources, 34*(7), 844-864. https://doi.org/10.1016/j.advwatres.2011.04.014

**Appendix A**

This appendix provides the data matching and prediction results for the observation point 3, 4, and 5 as introduced in subsection 6.5. In this case, only the hydraulic heads of the first 30 time-steps are available, as observation data and the hydraulic heads of the last 20 time-steps are utilized for testing the prediction from the estimated model parameters. The blue dashed lines in the figures indicate the beginning of prediction. It can be seen that the hydraulic heads from the updated model parameters can match the reference well in the prediction period.

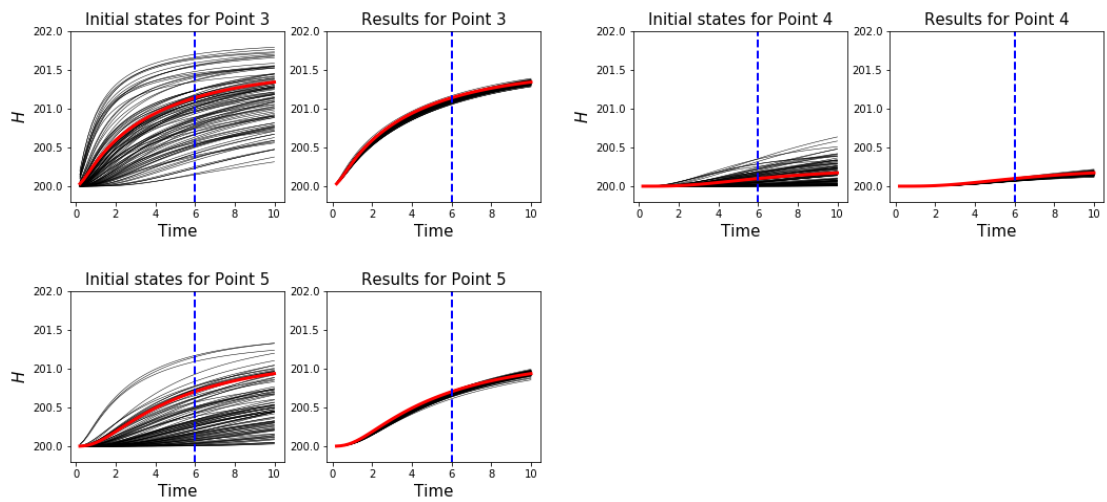

(a) Data matching and prediction results with the IES method

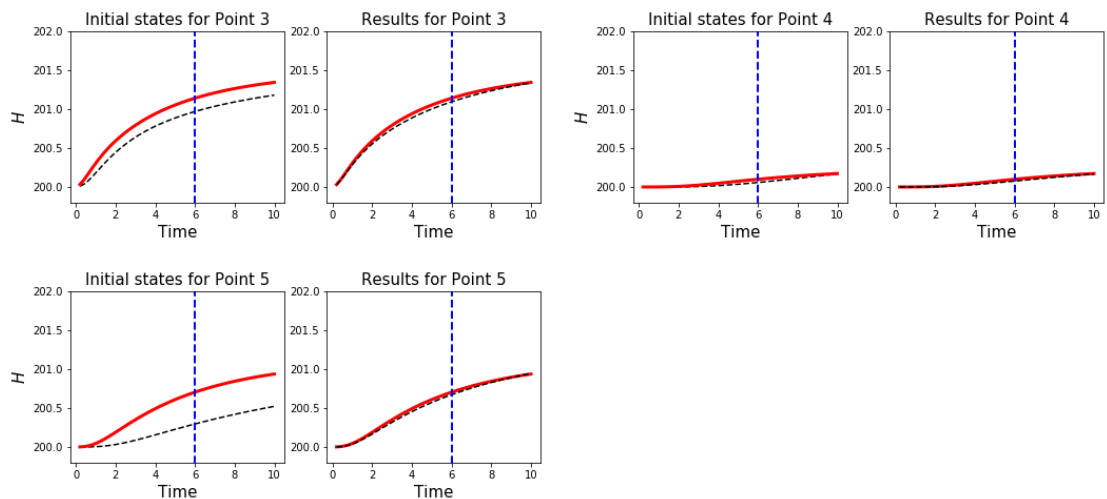

(b) Data matching and prediction results with the gradient method

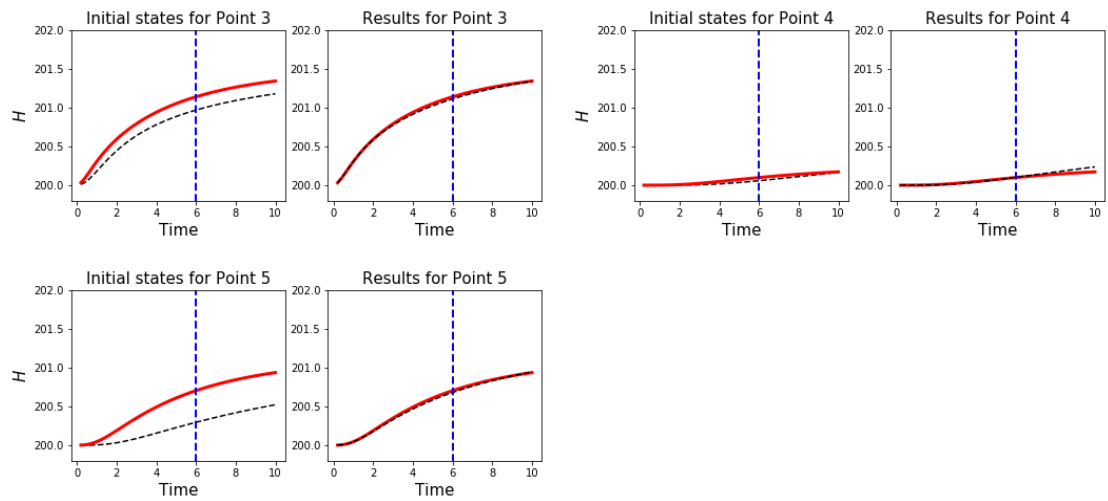

(c) Data matching and prediction results with the training method

**Figure A. 1.** Data matching and prediction results of points 3, 4, and 5 with different methods: IES method (a); gradient method (b); training method (c).